\documentclass[twoside,twocolumn,9pt]{article}
\usepackage{extsizes}
\usepackage[super,sort&compress,comma]{natbib} 
\usepackage[version=3]{mhchem}
\usepackage[left=1.5cm, right=1.5cm, top=1.785cm, bottom=2.0cm]{geometry}
\usepackage{balance}
\usepackage{mathptmx}
\usepackage{sectsty}
\usepackage{graphicx} 
\usepackage{lastpage}
\usepackage[format=plain,justification=justified,singlelinecheck=false,font={stretch=1.125,small,sf},labelfont=bf,labelsep=space]{caption}
\usepackage{float}
\usepackage{fancyhdr}
\usepackage{fnpos}
\usepackage[english]{babel}
\addto{\captionsenglish}{%
  
}
\usepackage{array}
\usepackage{droidsans}
\usepackage{charter}
\usepackage[T1]{fontenc}
\usepackage[usenames,dvipsnames]{xcolor}
\usepackage{setspace}
\usepackage[compact]{titlesec}
\usepackage{hyperref}

\usepackage{gensymb}
\usepackage{upgreek}
\usepackage{multirow}
\usepackage{colortbl}
\usepackage{float}

\usepackage{epstopdf}

\definecolor{cream}{RGB}{222,217,201}

\usepackage[normalem]{ulem} 

\begin{document}

\pagestyle{fancy}
\thispagestyle{plain}
\fancypagestyle{plain}{
\renewcommand{\headrulewidth}{0pt}
}

\makeFNbottom
\makeatletter
\renewcommand\LARGE{\@setfontsize\LARGE{15pt}{17}}
\renewcommand\Large{\@setfontsize\Large{12pt}{14}}
\renewcommand\large{\@setfontsize\large{10pt}{12}}
\renewcommand\footnotesize{\@setfontsize\footnotesize{7pt}{10}}
\makeatother

\renewcommand{\thefootnote}{\fnsymbol{footnote}}
\renewcommand\footnoterule{\vspace*{1pt}%
\color{cream}\hrule width 3.5in height 0.4pt \color{black}\vspace*{5pt}} 
\setcounter{secnumdepth}{5}

\makeatletter 
\renewcommand\@biblabel[1]{#1}            
\renewcommand\@makefntext[1]%
{\noindent\makebox[0pt][r]{\@thefnmark\,}#1}
\makeatother 
\renewcommand{\figurename}{\small{Fig.}~}
\sectionfont{\sffamily\Large}
\subsectionfont{\normalsize}
\subsubsectionfont{\bf}
\setstretch{1.125} 
\setlength{\skip\footins}{0.8cm}
\setlength{\footnotesep}{0.25cm}
\setlength{\jot}{10pt}
\titlespacing*{\section}{0pt}{4pt}{4pt}
\titlespacing*{\subsection}{0pt}{15pt}{1pt}

\fancyfoot{}
\fancyfoot[LO,RE]{\vspace{-7.1pt}\includegraphics[height=9pt]{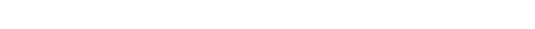}}
\fancyfoot[CO]{\vspace{-7.1pt}\hspace{13.2cm}\includegraphics{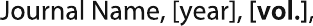}}
\fancyfoot[CE]{\vspace{-7.2pt}\hspace{-14.2cm}\includegraphics{head_foot/RF}}
\fancyfoot[RO]{\footnotesize{\sffamily{1--\pageref{LastPage} ~\textbar  \hspace{2pt}\thepage}}}
\fancyfoot[LE]{\footnotesize{\sffamily{\thepage~\textbar\hspace{3.45cm} 1--\pageref{LastPage}}}}
\fancyhead{}
\renewcommand{\headrulewidth}{0pt} 
\renewcommand{\footrulewidth}{0pt}
\setlength{\arrayrulewidth}{1pt}
\setlength{\columnsep}{6.5mm}
\setlength\bibsep{1pt}

\makeatletter 
\newlength{\figrulesep} 
\setlength{\figrulesep}{0.5\textfloatsep} 

\newcommand{\topfigrule}{\vspace*{-1pt}%
\noindent{\color{cream}\rule[-\figrulesep]{\columnwidth}{1.5pt}} }

\newcommand{\botfigrule}{\vspace*{-2pt}%
\noindent{\color{cream}\rule[\figrulesep]{\columnwidth}{1.5pt}} }

\newcommand{\dblfigrule}{\vspace*{-1pt}%
\noindent{\color{cream}\rule[-\figrulesep]{\textwidth}{1.5pt}} }

\makeatother

\twocolumn[
  \begin{@twocolumnfalse}
{\includegraphics[height=30pt]{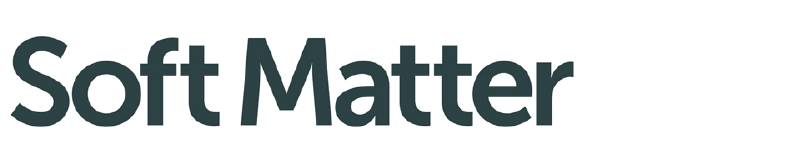}\hfill\raisebox{0pt}[0pt][0pt]{\includegraphics[height=55pt]{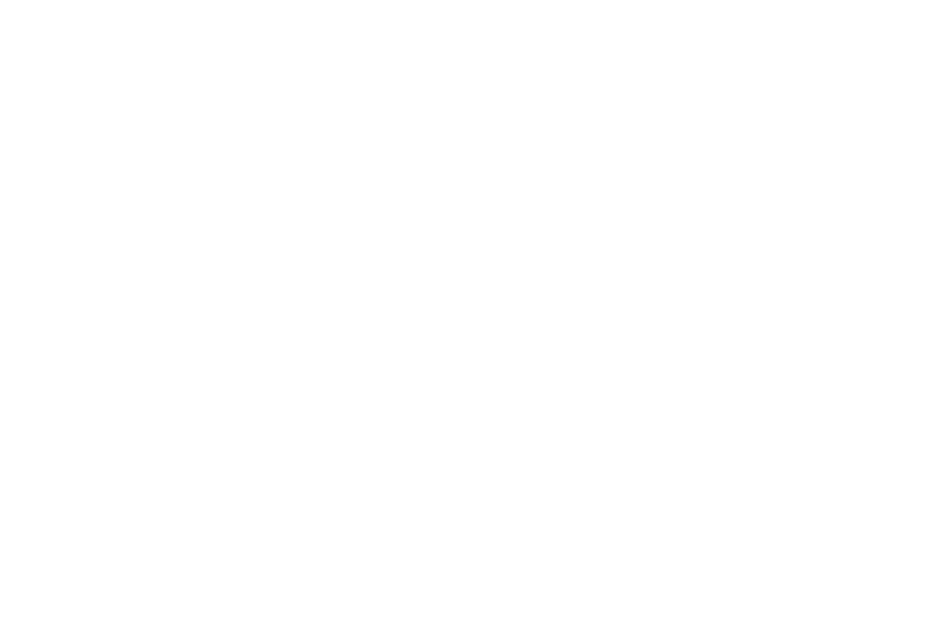}}\\[1ex]
\includegraphics[width=18.5cm]{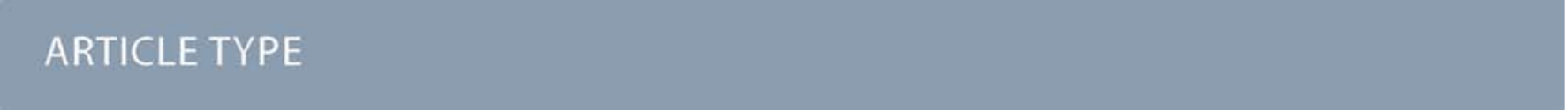}}\par
\vspace{1em}
\sffamily
\begin{tabular}{m{4.5cm} p{13.5cm} }

\includegraphics{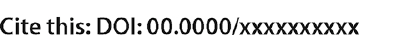} & \noindent\LARGE{\textbf{Robust fabrication of ultra-soft tunable PDMS microcapsules as a biomimetic model for red blood cells$^\dag$}} \\
\vspace{0.3cm} & \vspace{0.3cm} \\

 & \noindent\large{Qi Chen,\textit{$^{a, b}$} Naval Singh,\textit{$^{a, b}$} Kerstin Schirrmann,\textit{$^{a, b}$} Qi Zhou,\textit{$^{c}$} Igor Chernyavsky\textit{$^{d, e}$} and Anne Juel$^{\ast}$\textit{$^{a,b}$}} \\

\includegraphics{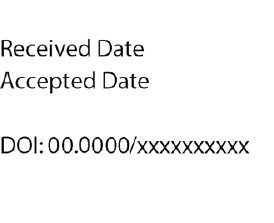} & \noindent\normalsize{Microcapsules with liquid cores encapsulated by thin membranes have many applications in science, medicine and industry. In this paper, we design a suspension of microcapsules which flow and deform like red blood cells (RBCs), as a valuable tool to investigate microhaemodynamics. A reconfigurable and easy-to-assemble 3D nested glass capillary device is used to robustly fabricate water-oil-water double emulsions which are then converted into spherical microcapsules with hyperelastic membranes by cross-linking the polydimethylsiloxane (PDMS) layer coating the droplets. The resulting capsules are monodisperse to within 1\% and can be made in a wide range of size and membrane thickness. We use osmosis to deflate by 36\% initially spherical capsules of diameter 350 $\upmu$m and a membrane thickness of 4\% of their radius, in order to match the reduced volume of biconcave RBCs. We compare the propagation of initially spherical and deflated capsules under constant volumetric flow in cylindrical capillaries of different confinements. We find that only deflated capsules deform broadly similarly to RBCs over a similar range of capillary numbers ($Ca$) -- the ratio of viscous to elastic forces. Similarly to the RBCs, the microcapsules transition from a symmetric 'parachute' to an asymmetric 'slipper'-like shape as $Ca$ increases within the physiological range, demonstrating intriguing confinement-dependent dynamics.  In addition to biomimetic RBC properties, high-throughput fabrication of tunable ultra-soft microcapsules could be further functionalized and find applications in other areas of science and engineering.
 
} \\




\end{tabular}

 \end{@twocolumnfalse} \vspace{0.6cm}
]

\renewcommand*\rmdefault{bch}\normalfont\upshape
\rmfamily
\section*{}
\vspace{-1cm}


\footnotetext{\textit{$^{a}$~Manchester Centre for Nonlinear Dynamics, The University of Manchester, Manchester, M13 9PL, UK. E-mail: anne.juel@manchester.ac.uk}} 

\footnotetext{\textit{$^{b}$~Department of Physics and Astronomy, The University of Manchester, Manchester, M13 9PL, UK. }}

\footnotetext{\textit{$^{c}$~School of Engineering, Institute for Multiscale Thermofluids, The University of Edinburgh, Edinburgh, EH9 3FB, UK. }}

\footnotetext{\textit{$^{d}$~Department of Mathematics, The University of Manchester, Manchester, M13 9PL, UK. }}

\footnotetext{\textit{$^{e}$~Maternal and Fetal Health Research Centre, School of Medical Sciences, The University of Manchester, Manchester, M13 9WL, UK. }}

\footnotetext{\dag~Electronic Supplementary Information (ESI) available: [details of any supplementary information available should be included here]. See DOI: 10.1039/cXsm00000x/}



\section{Introduction}\label{sec:introduction}

\begin{figure}[h!]
    \centering
    \includegraphics[width = 0.4 \textwidth]{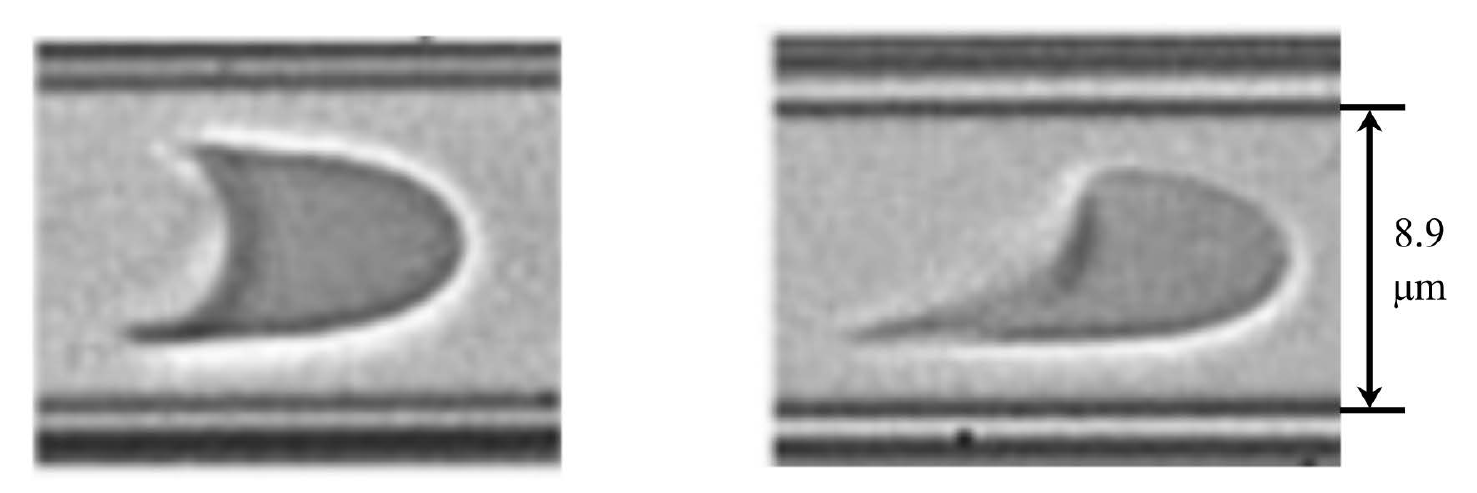}
    \caption{Symmetrical parachute-like (left) and symmetry-broken slipper-like (right) shapes of flowing RBCs at steady state in a cylindrical glass capillary tube \cite{Abkarian_2008}.}
    \label{fig:RBCs}
\end{figure}

Microcapsules with core-shell structures are widely used in industrial and biomedical domains, such as thermal energy storage \cite{Hao_2022}, enhanced oil recovery \cite{Fang_2019}, targeted drug delivery and controlled release \cite{Dinh_2020, Lee_2017}, and cell encapsulation and culture \cite{Mao_2016}. They also occur naturally in the form of, e.g., red blood cells (RBCs), bacteria and egg cells. Capsules that are used as carriers in confined media of complex geometry, e.g., enhanced oil recovery and biological delivery systems, have to be highly deformable and sufficiently robust to propagate efficiently and reach their targets. For example, the high deformability of RBCs enables them to flow through narrow pores and vessels of comparable size to deliver oxygen from the lungs to the rest of the body and carry carbon dioxide back to the lungs to be exhaled. Due to their high deformability, RBCs can adopt a variety of shapes in confined flow, including symmetrical parachute and symmetry-broken slipper-like shapes \cite{Abkarian_2008,Guckenberger_2018} (see Fig. \ref{fig:RBCs}) for sufficiently large capillary numbers $Ca$ -- a measure of the ratio of viscous to elastic forces acting on the capsule. Although the flow of isolated capsules (or RBCs) in confined vessels has been extensively studied \cite{Barth_s_Biesel_2016}, the fluid dynamics of capsule suspensions is yet to be fully addressed, including their propagation in complex media \cite{Zhou_2022_2} and rheological properties \cite{Beris_2021}. 

Recently, numerical simulations have been applied to investigate the dynamics of RBC suspensions in some complex geometries. However, they are usually limited to low haematocrit (or RBC volume fraction) and small flow domains because of their computational expense \cite{Zhou_2022_2}. The use of real RBCs in controlled experiments is also challenging because of their small size (typically less than 10 $\upmu$m), fragility, biological variability, rapid ageing and limited availability \cite{Carneiro_2021, Koch_2019}. Thus, various analogue models including capsules, elastic beads, vesicles and droplets have been developed to mimic flow behaviours of RBCs \cite{Zhou_2022,Rubio_2022,Sadek_2021}. Capsules have a core-membrane structure similar to RBCs and they typically exhibit larger deformations under flow than solid elastic beads \cite{Barth_s_Biesel_2016,Pinho_2020}. Although vesicles \cite{Misbah_2012} and droplets \cite{Bento_2018} can also deform considerably, they do not exhibit the shear elasticity of real RBCs \cite{Puthumana_2022}. Undesired coalescence of droplets further limits their use to dilute suspensions (equivalent to low haematocrit values). However, the geometrical and mechanical properties of microcapsules, such as membrane thickness and elastic modulus, need to be carefully chosen to access the deformations routinely observed in RBCs. Besides, to the best of our knowledge, only spherical objects have thus far been used in experiments to model RBCs, which adopt biconcave discoid shapes \cite{Wang_2012} at rest characterised by a reduced volume of 64\% of that of a sphere with the same surface area. The reduced volume of RBCs has been shown both experimentally and numerically to considerably affect their ability of passing through the confined microchannels \cite{Namvar_2020} and deformation in shear flow \cite{Zhu_2022}. The deformation and buckling of a spherical capsule as a function of reduced volume has been modelled \cite{Quilliet_2012} and validated with experiments on beach balls \cite{Coupier_2019}. However, characterisation of the motion and deformation of partially deflated capsules under shear flow is still lacking. Thus, there is a clear need for a robust approach to fabricate large populations of monodisperse microcapsules with ultra-thin soft membranes and a similar reduced volume to RBCs.

Droplet microfluidics is increasingly applied for capsule synthesis because this approach greatly simplifies the fabrication process by enabling precise control over the manufacturing conditions. The strategies typically encompass double emulsion templates, which consist of droplets encapsulated by an immiscible middle layer (drop-in-drop structure) and suspended in an outer liquid. The droplet coating is cured using either photo or thermally induced free-radical polymerisation \cite{do_Nascimento_2017, Hennequin_2009}, complexation reactions \cite{Zhang_2016} or solvent evaporation \cite{Kim_2011}. The membrane thickness is thus determined by the thickness of the coating layer. The axisymmetric glass capillary device firstly proposed by Utada et al. \cite{Utada_2005} is now most commonly used to produce double-emulsion droplets. This device is assembled with an array of nested glass capillaries that are pre-treated with silanes to achieve the required wettability distribution. Bandulasena \emph{et al.} \cite{Bandulasena_2019} and Levenstein \emph{et al.} \cite{Levenstein_2016} optimised the device fabrication method to be low-cost, time-saving and reconfigurable. This microfluidic method also offers flexibility over coating materials and thus membrane properties \cite{Chen_2011}.
Common materials used for capsule fabrication are UV-curable acrylates \cite{Hennequin_2009}, alginates \cite{Mou_2020} and proteins \cite{H_ner_2019, de_Loubens_2015}. However, these materials are either not sufficiently soft or exhibit rapid ageing, making them inappropriate for use in RBC analogues. In contrast, polydimethylsiloxane (PDMS) is a low-cost, transparent and flowable material which has been extensively characterised because of its ubiquitous use in microfabrication. It can be rapidly cross-linked under elevated temperature and exhibits very stable physical and chemical properties once cured, allowing the capsules to be stored for months. Besides, the mechanical properties of the membrane can be easily adjusted by mixing the PDMS base and the crosslinker in different ratios (see \S~\ref{sec:mechanical properties}).

The motion and deformation of single spherical capsules in confined flow have been extensively investigated both experimentally and numerically over the last two decades \cite{Barth_s_Biesel_2016}. Risso \emph{et al.} \cite{RISSO_2006} showed experimentally that the steady propagation of a capsule in a capillary tube is governed by the capillary number $Ca$ and the confinement parameter which is the ratio of capsule to tube diameter, and that capsule deformation is promoted by increase of either parameter. In a sufficiently confined geometry, the capsule extends into a barrel shape with spherical caps at both ends as $Ca$ increases. Beyond a critical value of $Ca$, the rear of the capsule buckles inward to form a parachute-like shape, similar to the deformation of RBCs. The transport of capsules also depends on the viscosity ratio between internal and external liquids ($\mu_{\rm int}$/$\mu_{\rm ext}$) and any pre-stress of the membrane. Although the viscosity ratio hardly affects steady capsule propagation, it can significantly influence transient behaviour by extending the time required for capsules with $\mu_{\rm int}$/$\mu_{\rm ext}> 1$ to reach a stable shape \cite{Diaz_2002}. Numerical simulations performed by Lefebvre \emph{et al.} \cite{Lefebvre_2007} show that pre-inflation of the capsule suppresses the appearance of the parachute shape to larger values of $Ca$. The elastic modulus of the capsules can be measured by compression testing, for millimetric capsules \cite{H_ner_2019}, and by atomic force microscopy or micropipette aspiration, for microcapsules, which requires skilled micro-manipulation \cite{Neubauer_2014}. In-flow measurements developed by Lefebvre \emph{et al.} \cite{Lefebvre_2008} are advantageous in that they circumvent size restrictions and enable high measurement throughput. However, they rely on a fit to numerical predictions from a membrane model based on a chosen constitutive behaviour. 
Individual capsules can also be trapped at the stagnation point of a microfluidic cross flow \cite{de_Loubens_2015} to ensure sufficiently large deformations albeit with the need for skilled user intervention. Finally, capsules and RBCs also exhibit rich dynamical behaviours beyond steady deformation, e.g., relaxation upon exit from a single channel constriction \cite{Rorai_2015}, or in complex channel geometries, such as T-junctions or networks \cite{H_ner_2019,Merlo_2022}, where they transiently adopt highly deformed slipper-like shapes. 


In this paper, we establish a robust methodology to manufacture large populations of highly monodisperse and stable microcapsules, which are partially deflated by osmosis to match the reduced volume of RBCs. The fabrication methods are described in \S~\ref{sec:cap fabrication}, while characterisation of capsule properties including size, membrane thickness, elastic modulus and deflation are discussed in \S~\ref{sec:Geo. and Mech.}. In \S~\ref{sec:capillary flow} we compare the steady flow behaviour of spherical and partially deflated microcapsules in capillary tubes as a function of the viscous-elastic capillary number $Ca$ and geometric confinement. Conclusions and outlook for using partially deflated microcapsules as RBC analogues are given in \S~\ref{Conc}. 

\section{Capsule fabrication}\label{sec:cap fabrication}

\begin{figure*}[h!]
    \centering
    \includegraphics[width=1.0\textwidth]{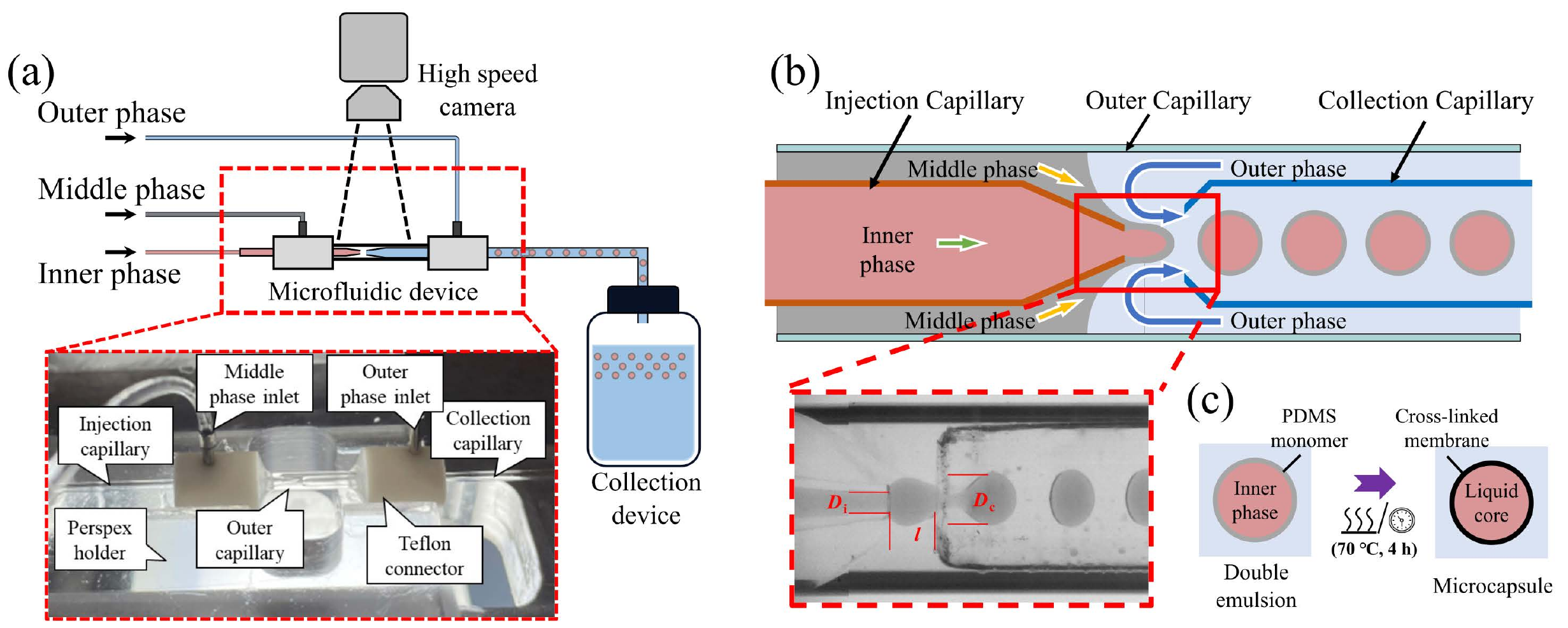}
    \caption{(a) Schematic diagram of the experimental setup for capsule fabrication. The inset shows the labelled image of the 3D nested glass capillary device mounted in a perspex holder. (b) Schematic illustration of the capillary device to make W/O/W double emulsions. The optical microscope image shows the formation of double emulsion droplets at typical dripping flow regime ($Q_{\rm i}$ = 24 $\upmu$L/min, $Q_{\rm m}$ = 3 $\upmu$L/min and $Q_{\rm o}$ = 150 $\upmu$L/min for the inner, middle and outer phases respectively). (c) Schematic diagram of microcapsule cured from double emulsion templates.}
    \label{fig:Exp. Setup}
\end{figure*}

We fabricated capsules by generating a train of coated droplets within a microfluidic nested-capillary device and collected the droplets by letting them fall from the end of a collection capillary bent at 90$\degree$ into a glass bottle, where they cured at rest; see Fig. \ref{fig:Exp. Setup}(a). The double-emulsion generation was monitored in top-view with a monochrome CMOS high-speed camera (PCO 1200hs) fitted with long-distance magnifying optics, which consisted of a zoom lens (Navitar, 12× Zoom Lens System) coupled to a 5× microscope objective (Mitutoyo, M Plan APO). The maximum combined magnification therefore reaches 60×, and images were recorded with a maximum frame rate of 500 frames per second (fps). The microfluidic device was backlit with uniform, diffuse illumination of adjustable brightness provided by a custom-made white LED light box.

The nested glass capillary device shown in Fig. \ref{fig:Exp. Setup}(a, b) comprised three glass capillaries of circular cross-section: injection and collection capillaries (O.D. = 1.0 mm, I.D. = 0.58 mm, World Precision Instruments) and an outer capillary (O.D. = 2.0 mm, I.D. = 1.8 mm, S Murray {\&} Co, UK). The outer capillary was snapped into two pieces of approximately 2 cm length after scoring its midpoint with a ceramic tile (Sutter Instrument). The injection capillary was tapered with a micropipette puller (P-97, Sutter Instrument) and its tip was cut off and polished using the tile to a diameter of $D_{\rm i}$ = 120 $\upmu$m. A gas torch flame was used to melt one end of the collection capillary to form a constricted nozzle-like structure. We treated several capillaries and selected those with a nozzle diameter $D_{\rm c} = 350 \pm 20$ $\upmu$m to use in the nested capillary device. The 90$\degree$ bend of the collection capillary was applied to the initially straight capillary held in a horizontal position, by heating its midpoint under the gas torch flame until the heat-softened region allowed the end of the capillary to drop spontaneously under gravity.  

We used a mixture of water, glycerol (Sigma Aldrich) and 2.0 wt\% PVA (Polyvinyl alcohol, partially hydrolysed, MW approx. 30000, Sigma-Aldrich) for both inner and outer phases, where water and glycerol were mixed in 36:64 by volume. Carminic acid (Sigma-Aldrich) was added at 0.1 wt\% to the inner phase to dye the capsules red. The middle phase was PDMS (Sylgard 184, Dow Corning) which could be mixed with different ratios of base to crosslinker to vary the elastic properties of the cured material. The mixture was degassed in a vacuum chamber for 20 minutes to extract dissolved air before transferring it to a 1 mm syringe (Injekt-F) to be injected with a syringe pump. Because the PDMS progressively cured over time, thus increasing its viscosity, the mixture had to be used within 3 hours to ensure stable droplet formation with consistent size and shape. To achieve water-oil-water (W/O/W) double emulsions, the injection capillary had to be chemically treated with Sigmacote (Sigma-Aldrich), a solution of chlorinated organopolysiloxane dissolved in heptane which adsorbs to the glass surface to form a hydrophobic coating. The collection and outer capillaries were treated with oxygen plasma (HPT-100, Henniker Plasma) at 100\% power for 3 minutes to enhance their hydrophilicity. 

Immediately after plasma treatment, the injection and collection capillaries were co-axially aligned inside the larger outer capillary aided by the magnified image from the camera. The outer capillary was held between two custom-made Teflon connectors (see the inset of Fig. \ref{fig:Exp. Setup}(a)) which were accurately drilled through along their central axis; see Fig. S1 in ESI for the detailed design. The injection and collection capillaries were each pushed through a connector so that the tapered end of the injection capillary was separated from the nozzle of the collection capillary by a distance $l \simeq$ 200 $\upmu$m, as shown in Fig. \ref{fig:Exp. Setup}(b). Fluid inlet tubes were connected at this stage and all joints were sealed and fixed with a UV curable glue consisting of a 70:30 (v/v) mixture of pentaerythriol triacrylate (PETA, Sigma-Aldrich) and Tri(propyleneglycol) diacrylate (TRPGDA, Sigma-Aldrich) with 5\% of 1-Hydroxycyclohexylphenyl ketone (Sigma-Aldrich) as the photoinitiator. This glue solidified within a few seconds under UV irradiation, which enabled rapid assembly of the device. The entire device was positioned inside a well milled Perspex sheet, with an observation window cut out to enable imaging of the double emulsion formation. The entire design was focused on enabling rapid and reproducible dismantling and reassembly of the device, which typically took 20 minutes if all accessories were readily available.

To generate the double emulsion, the inner and middle phases were injected with constant volumetric rates ($Q_{\rm i}$ and $Q_{\rm m}$) into the injection capillary and the gap between the injection and outer capillaries through a tube into the leftmost connector, respectively. The outer phase was injected at $Q_{\rm o}$ through the gap between the collection and outer capillaries through a tube into the rightmost connector (see Fig. \ref{fig:Exp. Setup}(a, b)). The inner and outer phases were injected under flow rate control using a pressure controller (Elveflow Mk3+ 0--2~Bar, Elvesys) coupled to in-line Mass Flow Sensors (MFS, 0--80 $\upmu$L/min, Elvesys). The middle phase was injected using a syringe pump (KD Scientific, Model 210) to avoid the occlusion of flow sensors because of the eventual curing of the middle phase. 

The three phases came into contact in the region of length $l$ between the injection and the collection capillaries (Fig. \ref{fig:Exp. Setup}(b)), where the inner phase coated by the middle phase broke into double emulsion droplets suspended in the outer phase. In order to achieve the interface shapes necessary for the stable generation of coated droplets, the fluids had to be introduced in strict order. We first flowed the outer phase into the system and let it wet the entire device before introducing the middle phase. A stable cone-like interface was formed owing to the hydrophobic injection capillary and the hydrophilic outer capillary, resulting in the formation of a regular train of middle phase droplets. Finally, the inner phase liquid was introduced in order for it to be completely surrounded by the middle phase upon exit of the injection capillary, and the train of middle-phase droplets was thus replaced by droplets of inner phase encapsulated by the middle phase. The flow rate for each phase was adjusted to retain a stable dripping regime, as shown in Fig. \ref{fig:Exp. Setup}(b). This capillary device could make in excess of one hundred capsules per minute under optimal operating conditions. Video 1 in ESI shows an example of double-emulsion generation under such optimal conditions. Sub-optimal conditions are shown in Video 2, where mixing of the inner and outer phases occurs due to the misalignment of the injection and collection capillaries; in Video 3 the interface breaks up due to defective hydrophobic treatment on the injection capillary; and in Video 4 an irregular interface forms due to insufficient plasma treatment on the outer and collection capillaries.


We typically generated populations of tens of thousands of coated droplets to be cured. The bottle in which they were collected contained a buffer layer of the suspending liquid (outer phase) to avoid direct contact with the bottom wall upon impact. After collection, the bottle was placed in an oven at 70$\,^{\circ}\mathrm{C}$ for 4 hours to cure the PDMS coating and convert the droplets into capsules, as illustrated in Fig. \ref{fig:Exp. Setup}(c). Curing the coating while the droplets were at rest ensured uniform membranes and avoided off-centre placement of the core \cite{Hennequin_2009}, in contrast with in-situ UV-polymerisation where moving droplets were irradiated in the outlet capillary while subject to shear forces. Shrinkage of the PDMS coating by $< 1.5\%$ during curing resulted in a slight extensional pre-stress being applied to the membrane. Although physical and chemical properties remained stable over periods of months, regular stirring of the suspension was required to avoid capsules sticking together due to the drainage of the lubrication layer separating them.

\section{Capsule characterisation}\label{sec:Geo. and Mech.}
\subsection{Geometrical properties}\label{subsec:geometrical properties}

\begin{figure}[h!]
    \centering
    \includegraphics[width=0.45\textwidth]{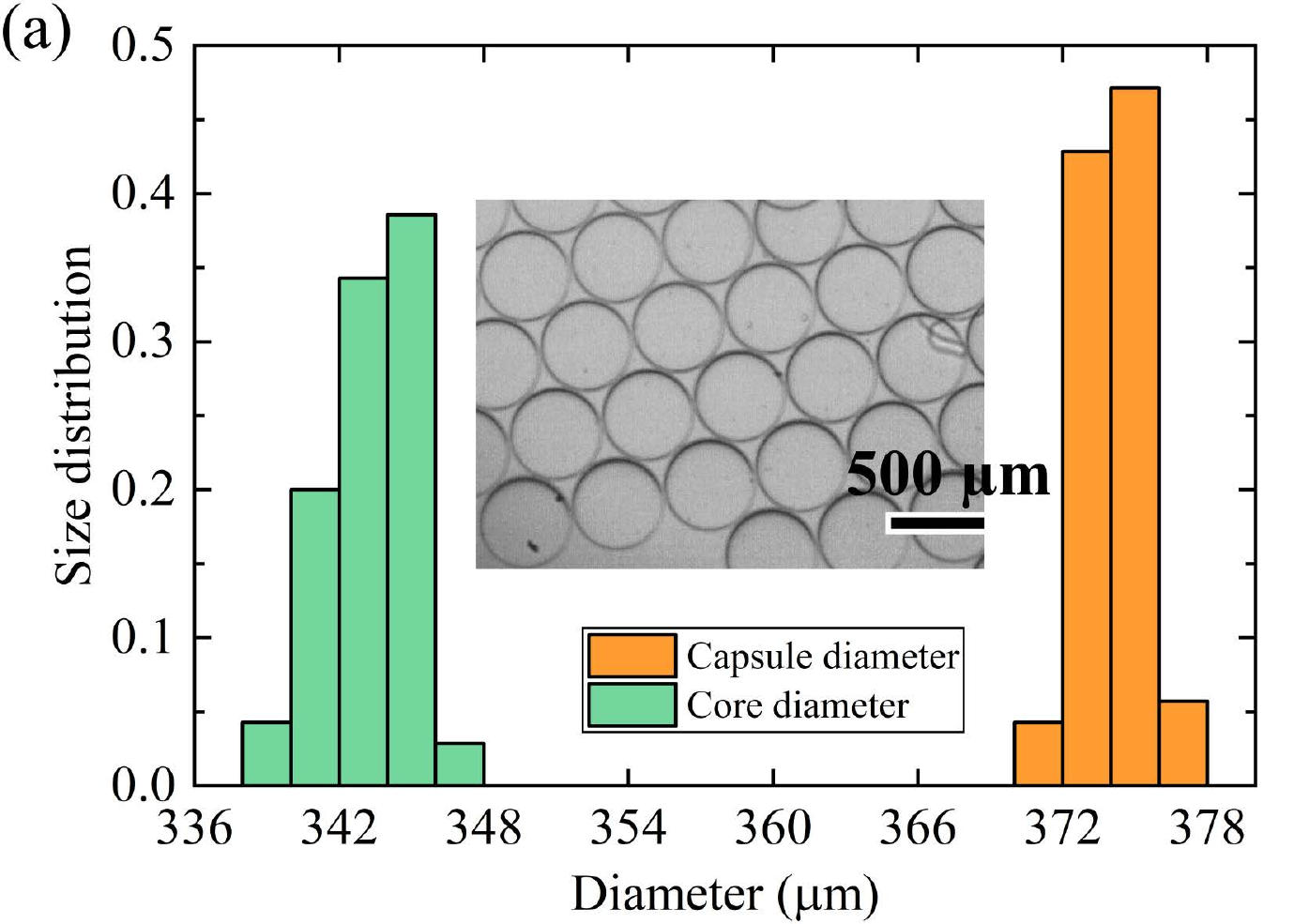}
    \includegraphics[width=0.45\textwidth]{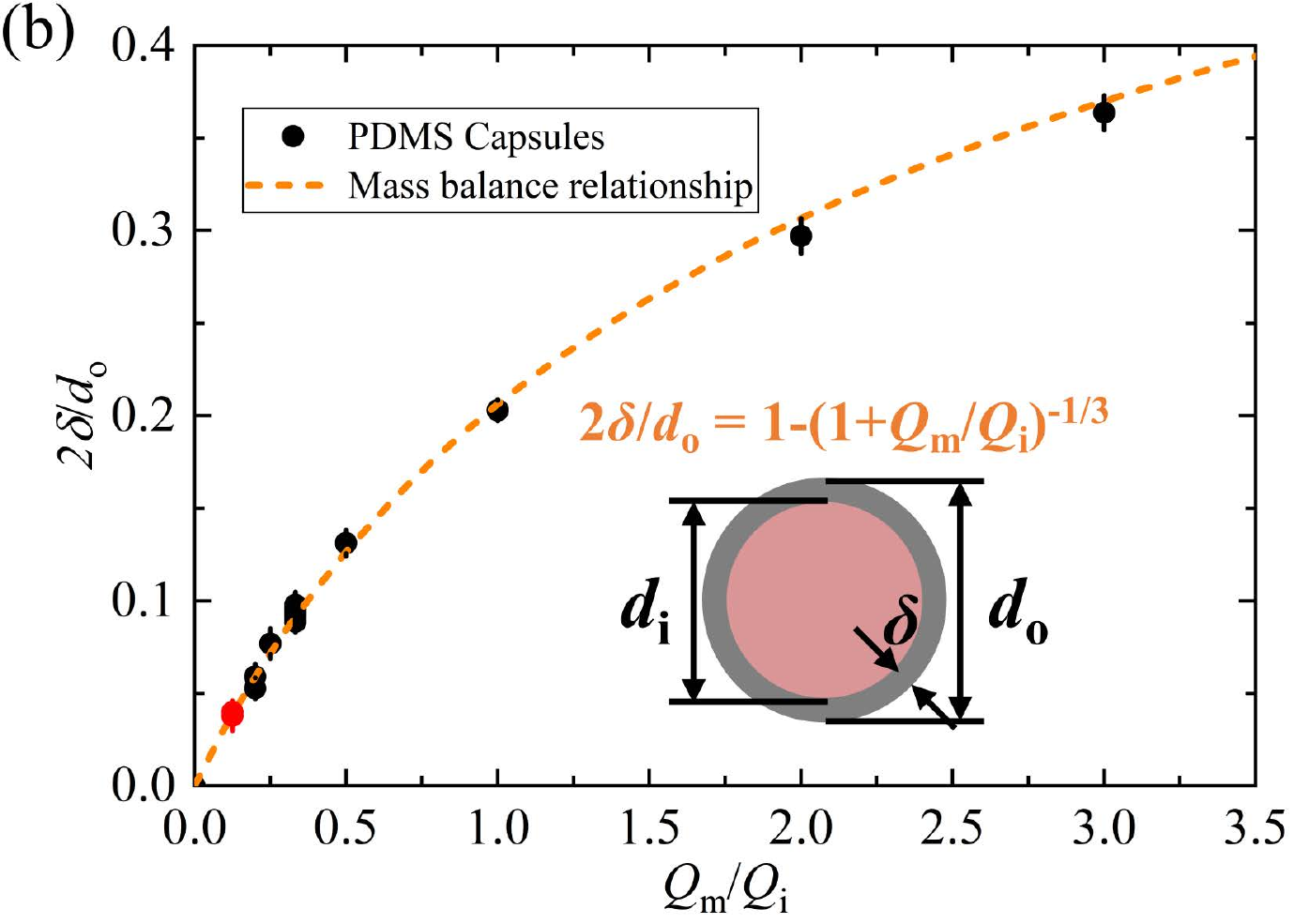}
    \includegraphics[width=0.45\textwidth]{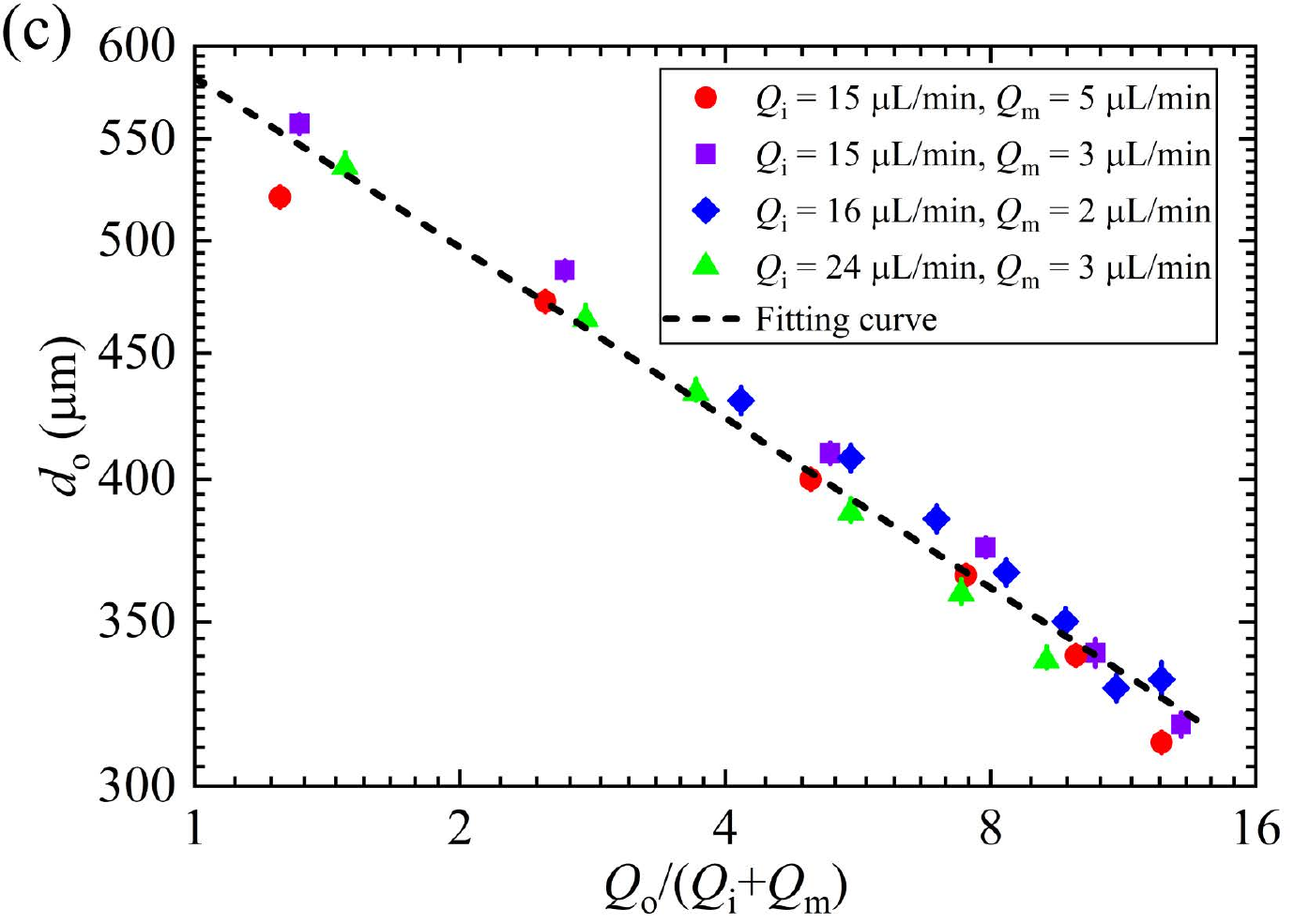}
    \caption{(a) Capsule diameter distribution determined for 70 PDMS capsules fabricated at $Q_{\rm i}$ = 35 $\upmu$L/min, $Q_{\rm m}$ = 10 $\upmu$L/min and $Q_{\rm o}$ = 300 $\upmu$L/min. (b) Relative membrane thickness 2$\delta$/$d \rm _o$ as a function of the middle to inner phase flow rate ratio $Q \rm _m/Q \rm _i$. The errors are the standard deviation of at least 50 capsule samples. Dashed line indicates the theoretical value  predicted by Eq. \ref{eq:membrane thickness} according to the mass conservation. The dots marked red indicate our red blood cell models. (c) Power-law fitting of the capsule diameter ($d_{\rm o}$) as a function of the flow-rate ratio $Q \rm _o$/($Q \rm _i+Q \rm _m$). The relative membrane thickness is keeping constant while changing the outer phase flow rates: red circles - 9.14\%, purple squares - 5.90\%, blue diamonds and green triangles - 3.85\%.}
    \label{fig:characterisation}
\end{figure}

Healthy human RBCs typically exhibit low size variability with typical diameters of 7--8 $\upmu$m, and thus any RBC analogue must also be highly monodisperse. Accurate control of capsule size and membrane thickness is also important to ensure reproducible experiments and meaningful upscaling from RBCs to capsules. 
The histogram in Fig. \ref{fig:characterisation}(a) shows the distribution of outer and inner capsule diameters measured from a sample of 70 capsules manufactured with $Q_{\rm i}$ = 35 $\upmu$L/min, $Q_{\rm m}$ = 10 $\upmu$L/min and $Q_{\rm o}$ = 300 $\upmu$L/min. The mean outer and inner diameters are $d_{\rm o}=373.7\pm1.8$ $\upmu$m and $d_{\rm i}= 343.9\pm 1.4$ $\upmu$m, respectively, with very small standard deviations of 0.4\% and 0.5\% of the mean values, respectively.

The capsule fabrication method also enabled us to customise the suspension by applying different flow rate combinations during double emulsion generation. The steady-state generation of double emulsions means that $Q_{\rm m}$ and $Q_{\rm i}$ contribute uniformly to the capsule's inner core and coating layer volumes, respectively. Thus, mass conservation, based on an incompressible spherical capsule, predicts the ratio of membrane thickness to the outer capsule radius $2\delta/d_{\rm o}$ that depends only on the ratio of the middle and inner phase flow rates $Q_{\rm m}/Q_{\rm i}$ as
\begin{equation}
    \frac{2\delta}{d_{\rm o}} = 1 - \left(1 + \frac{Q_{\rm m}}{Q_{\rm i}}\right)^{-\frac{1}{3}}.
    \label{eq:membrane thickness}
\end{equation}
Fig.~\ref{fig:characterisation}(b) shows that Eq.~(\ref{eq:membrane thickness}) (the continuous dashed line in Fig.~\ref{fig:characterisation}) accurately predicts the relative membrane thickness measured experimentally in the range from approximately 4\% to 37\% (solid circles). Rachik \emph{et al.} \cite{Rachik_2006} compared experimental characterisations of serum albumin–alginate capsules with model predictions to find that the thin shell assumption is only valid for relative membrane thickness of less than 5\% beyond which it is necessary to consider bulk elastic effects. Thus, we kept the relative membrane thickness of our capsules at 4\% to simplify the problem, as indicated with a red symbol in Fig. \ref{fig:characterisation}(b). 

We also found that the variation of the outer phase flow rate $Q_{\rm o}$ does not affect the relative membrane thickness but will change the capsule size, which provides an opportunity to customise the capsule diameter without perturbing the relative membrane thickness. Fig. \ref{fig:characterisation}(c) shows that the capsule size decreases with increasing $Q_{\rm o}$, while $Q_{\rm m}$ and $Q_{\rm i}$ are kept fixed to maintain a constant relative membrane thickness. We were able to vary the outer capsule diameter in the range 300 $\upmu$m\textless$d{\rm _o}$\textless550 $\upmu$m by varying $Q_{\rm o}$. To extend the range of capsule diameters further requires the adjustment of the collection capillary diameter $D_{\rm c}$, as discussed by Michelon \emph{et al.} \cite{Michelon_2020}. Vladisavljevi{\'{c}} \emph{et al.} \cite{Vladisavljevi__2012} analysed the double emulsion generation in a similar device based on mass conservation and found that the droplet diameter has a power-law dependence with an index of $-1/3$ to the droplet generation frequency that is directly determined by $Q_{\rm o}$. So, we can also get a power-law relation between the capsule diameter $d_{\rm o}$ and $Q_{\rm o}$ as 
\begin{equation}
    d_{\rm o} = a \left(\frac{Q_{\rm o}}{Q_{\rm i}+Q_{\rm m}}\right)^b.
    \label{eq:Capsule size}
\end{equation}
Because the generation frequency itself is non-linear in $Q_{\rm o}$, the index $b$ in this equation is different from $-1/3$ depending on the dimensions of the device and physical properties of the liquids. Michelon \emph{et al.} \cite{Michelon_2020} reported an index of $-0.3$ in their experiments. Fig.~\ref{fig:characterisation}(c) shows that our capsule diameter also agrees well with Eq. \ref{eq:Capsule size} (the continuous dashed line), giving $a = 583 \pm 7$ $\upmu$m and $b = -0.23 \pm 0.01$. For our scaled-up RBC model, the capsule size is kept at approximately 350 $\upmu$m.

\subsection{Mechanical properties}\label{sec:mechanical properties}

\begin{figure}[h!]
    \centering
    \includegraphics [width=0.45\textwidth] {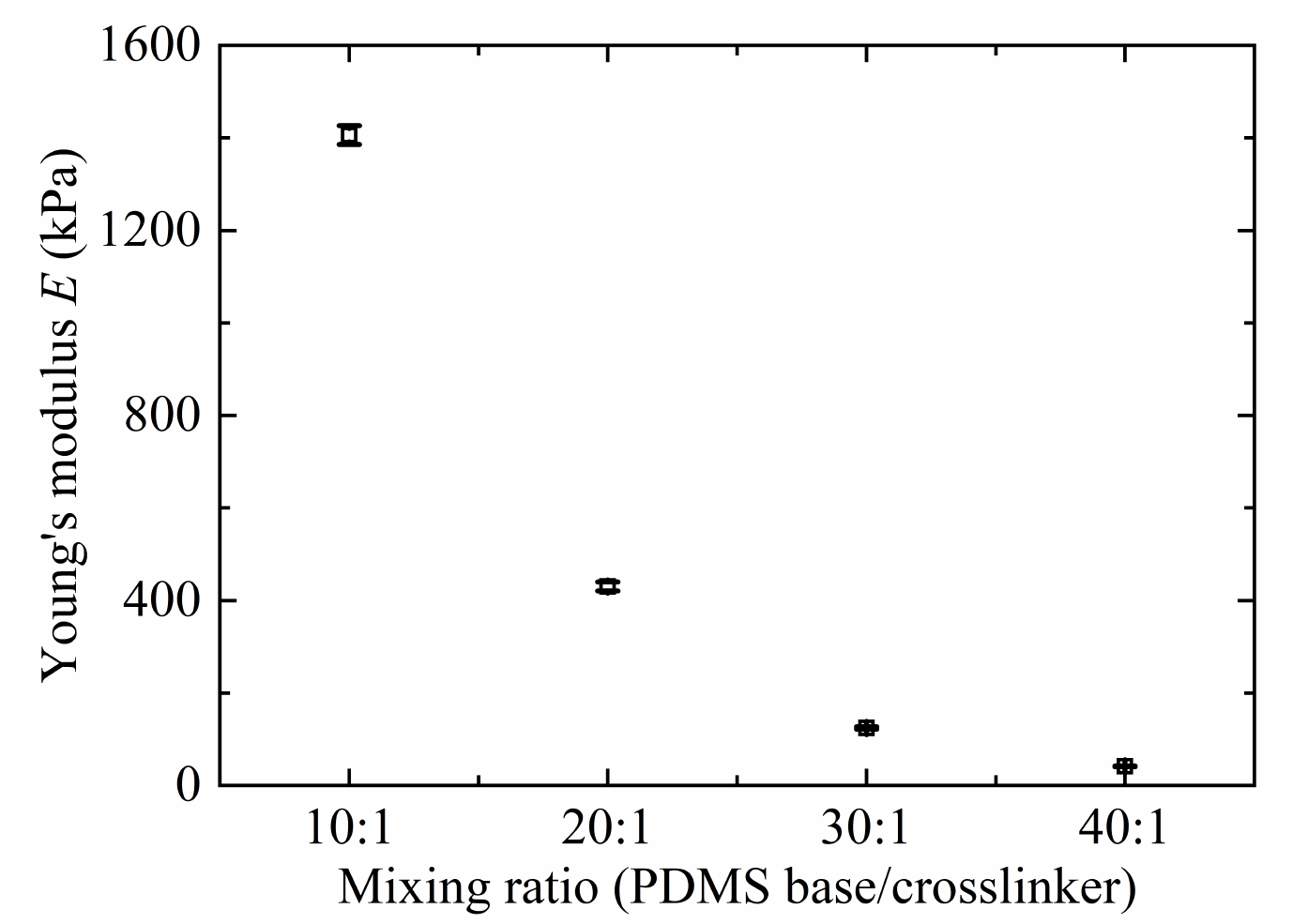}
    \caption{Young's modulus as a function of the mixing ratio of PDMS base to crosslinker predicted by Mooney--Rivlin model \cite{willshaw2012pattern}. Each point corresponds to the mean value of five replicate measurements, and the error bars come from errors of the fitting parameter.}
    \label{fig:young's modulus}
\end{figure}

To characterise the elastic properties of the membrane, we measured Young's modulus $E$ by compression testing of a cylindrical sample and deduced the shear modulus theoretically in the limit of linear elasticity. Simply put, we measured the engineering stress as a function of the engineering strain, and the data is accurately captured by the Mooney-Rivlin model \cite{willshaw2012pattern}, from which we deduced $E$ in the limit of vanishing strain. The details of this method are discussed further in S3 of ESI. Experimental measurement of Young's modulus as a function of the mass ratio of the PDMS base to the crosslinker are shown in Fig. \ref{fig:young's modulus}. Increasing the mixing ratio from 10:1 to 40:1 decreases Young's modulus by more than an order of magnitude from 1.4 MPa to 42 kPa. For isotropic polymeric materials, the bulk shear modulus $G$ is related to Young's modulus $E$ via Poisson's ratio $\nu$,
\begin{equation}
    G = \frac{E}{2(1 + \nu)},
    \label{eq:Bulk shear modulus}
\end{equation}
where $\nu \simeq 0.5$ for PMDS which is assumed incompressible. The capsule membrane is treated as a thin sheet of a 3D homogeneous hyperelastic material with thickness $\delta$, and the thin shell approximation is applied on the membrane mid-surface to compute the 3D effects \cite{Chapelle_2011}. Then, the surface shear modulus $G \rm _s$ (N/m) is calculated based on the bulk shear modulus $G$ by 
\begin{equation}
    G \rm _s = G\delta, \label{eq:surface shear modulus}
\end{equation}
and the bending modulus $\kappa$ is \cite{Barth_s_Biesel_2016}:
\begin{equation}
    \kappa = \frac{G\delta^3}{6(1 - \nu)}.
\end{equation}
For 350 $\upmu$m capsules with a membrane thickness of 4\% of the radius, the surface shear elastic modulus $G \rm _s=0.098$~N/m and the bending modulus $\kappa=1.6\times 10^{-12}$ N$\cdot$m. We also define the capillary number ($Ca$) based on the viscosity of the suspending fluid $\mu_{\rm ext}$ and the surface shear modulus $G_{\rm s}$ as
\begin{equation}
    Ca = \frac{\mu_{\rm ext} \bar{u}}{G_{\rm s}},
    \label{eq:Ca}
\end{equation}
where $\bar{u}$ is mean velocity of flow. 

\subsection{Controlled capsule deflation}\label{sec:cap deflation}

\begin{figure}[h!]
    \centering
    \includegraphics[width=0.45\textwidth]{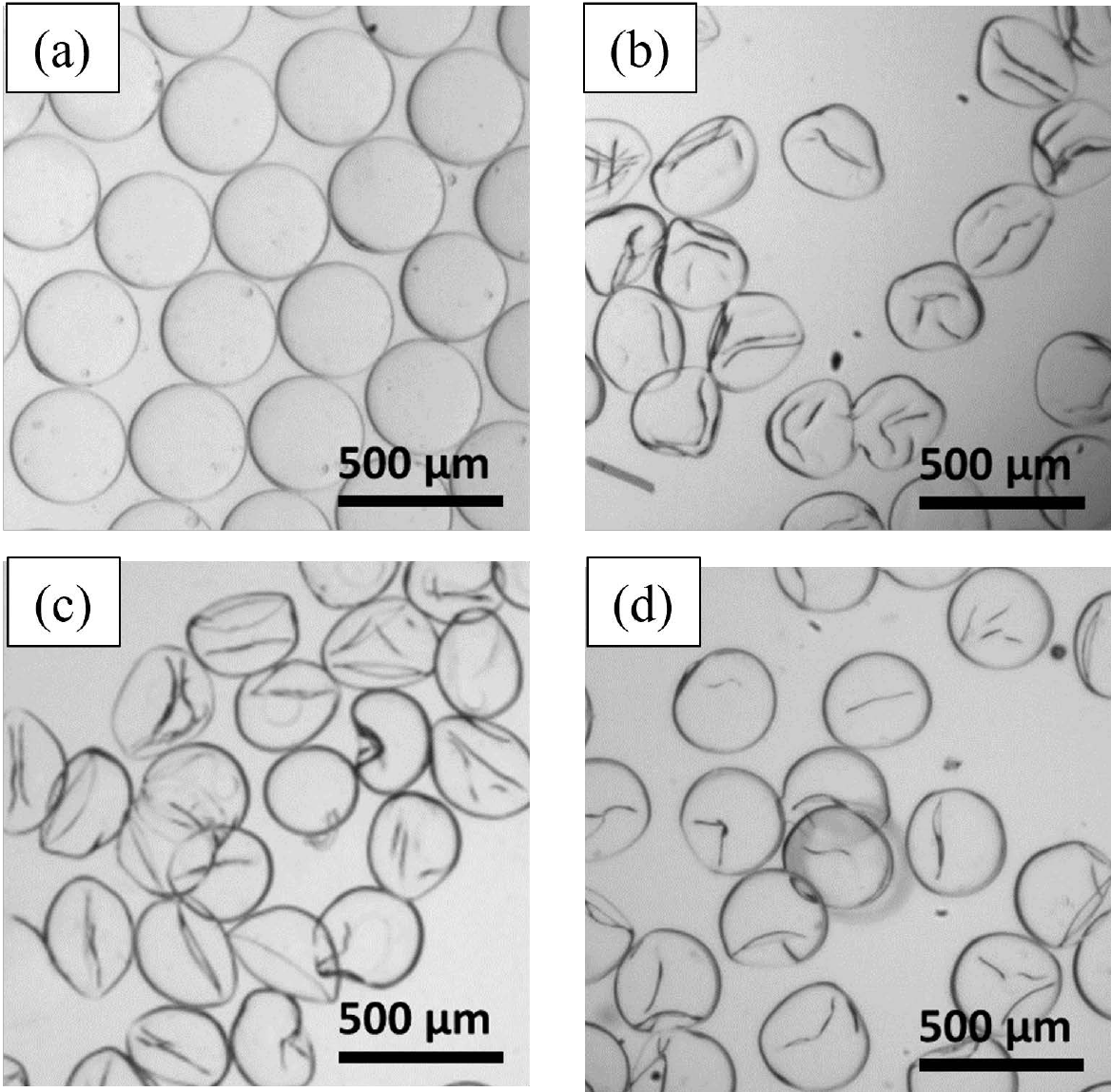}
    \caption{Capsules with different levels of deflation (characterised by the reduced volume ratio $\alpha$): (a) Spherical capsules after fabrication, $\alpha$ = 1; (b) Capsules deflated by 80\%, $\alpha$ = 0.2; (c) Capsules deflated by 60\%, $\alpha$ = 0.4; (d) Capsules deflated by 36\%, $\alpha$ = 0.64, corresponding to the RBC value. The capsules are made of PDMS and cross-linker mixed in 40:1, and their initial diameters $d_{\rm o}$ are 350 $\upmu$m with a relative membrane thickness $2\delta/d_{\rm o} = 4\%$.}
    \label{fig:capsule deflation}
\end{figure}

With a typical surface area $A \simeq 140 \,\upmu$m$^2$ and volume $V \simeq 100 \, \upmu$m$^3$, RBCs have a large surface area to volume ratio of $A/V \simeq 1.56$ times that of an equivalent sphere with the same surface area because of their biconcave shape at rest. A sphere with the same surface area as the RBC would have a radius $R_{\rm e}$ such that $4 \pi R_{\rm e}^2 = A$, which yields $R_{\rm e} \simeq 3.34\, \upmu$m. Thus, the reduced volume of RBCs is $\alpha = V/V_{\rm e} = 0.64$, where $V_{\rm e}$ is the volume of the equivalent sphere  \cite{Zhou_2022_2}. 
This reduced volume can be matched in our analogue capsule model by deflating spherical capsules by 36\%. We achieved capsule deflation through osmosis because the PDMS membrane is permeable to water but not to glycerol and thus, the osmotic pressure generated by a larger concentration of water inside the capsule core than in the suspending buffer solution in the collection bottle drives water across the membrane and out of the capsule until concentrations equilibrate. We started from a population of spherical capsules, a sample of which is visualised on a microscope glass slide in Fig.~\ref{fig:capsule deflation}(a). Although both the inner phase, forming the core of the capsule, and the buffer solution in the bottle had the same mixing ratio of water to glycerol, the slightly lighter PDMS membrane meant that capsules spontaneously rose to the upper surface of the buffer solution in the bottle over a period of a few hours. Hence, we could easily remove the suspending liquid with a syringe and replace it with pure glycerol. The capsules were then carefully stirred into the pure glycerol with a glass bar taking utmost care to avoid trapping any air bubbles. The osmotic pressure generated by the difference in water concentration between the capsule core and pure glycerol in the bottle drove water across the membrane until the capsule core was mostly depleted of water because of the large volume of glycerol in the bottle. This process took about 30 minutes. The suspension was then left to rest for a few hours to allow the capsules to rise to the top surface again. We then repeated the process at least three times to ensure that all water in the capsule core had been removed.


By adjusting the mixing ratio of water to glycerol in the capsule core, we produced microcapsules with different levels of deflation. Fig. \ref{fig:capsule deflation}(b--d) show deflated capsules with volumes reduced by 80\%, 60\% and 36\%, respectively. The capsules with 36\% volume reduced match the reduced volume $\alpha$ of real RBCs. A 36:64 by volume solution of water and glycerol was used for the inner and outer phases to generate double emulsion required for these capsules. 

\section{Steady flow and deformation of spherical and deflated capsules in capillaries}\label{sec:capillary flow}

We now compare the steady flow and deformation of spherical and deflated capsules (with a reduced volume of 0.64 similar to RBCs) in cylindrical glass capillary tubes of inner diameters $D =0.4,\, 0.3, \, 0.2$~mm.  The experiments were performed by injecting a very dilute suspension of capsules into a capillary tube at constant volumetric rate using a syringe pump (KD Scientific, Model 210), which ensured that the capsules were sufficiently separated to avoid measurable interaction within the tube. We used the high speed camera fitted with long-distance magnifying optics (Navitar 12× Zoom Lens System coupled with a 10× microscope objective) to capture images of the capsule shapes. Images were recorded at a frame rate between 125 to 1500 fps (depending on the flow rate) with an exposure time of 1/3000 second. The capillary tube was backlit with a powerful cold-light source (Karl Storz - Xenon Nova 300).

\begin{table}[h!]
\centering
\caption{Comparison between key capsule parameters. All measurements were taken at $(20 \pm 1)^{\circ}$C. \label{Table:capsule parameters}}
\begin{tabular}{l c c} 
\hline
Parameters & \begin{tabular}[c]{@{}c@{}}Spherical \\ capsules\end{tabular} &  \begin{tabular}[c]{@{}c@{}}36\% deflated \\ capsules\end{tabular}\\ 
\hline
Reduced volume ratio $\alpha$                    & 1       & 0.64  \\
Effective diameter ($d_{\rm eff}$, $\upmu$m)               & 350     & 302   \\
\begin{tabular}[l]{@{}lc@{}} Relative membrane thickness \\ ($2\delta/d_{\rm o}$)\end{tabular}          & 4\%     & 4\%   \\
\begin{tabular}[l]{@{}lc@{}} Shear elastic modulus \\ ($G_{\rm s}$, N/m) \end{tabular}         & 0.098   & 0.098 \\
External viscosity ($\mu_{\rm ext}$, Pa$\cdot$s) & 0.353   & 1.55 \\
Viscosity ratio ($\mu_{\rm int}/\mu_{\rm ext}$)  & 1.00    & 1.00 \\
\hline
\end{tabular}
\end{table}

\begin{figure*}[h!]
    \centering
    \includegraphics[width=1.0\textwidth]{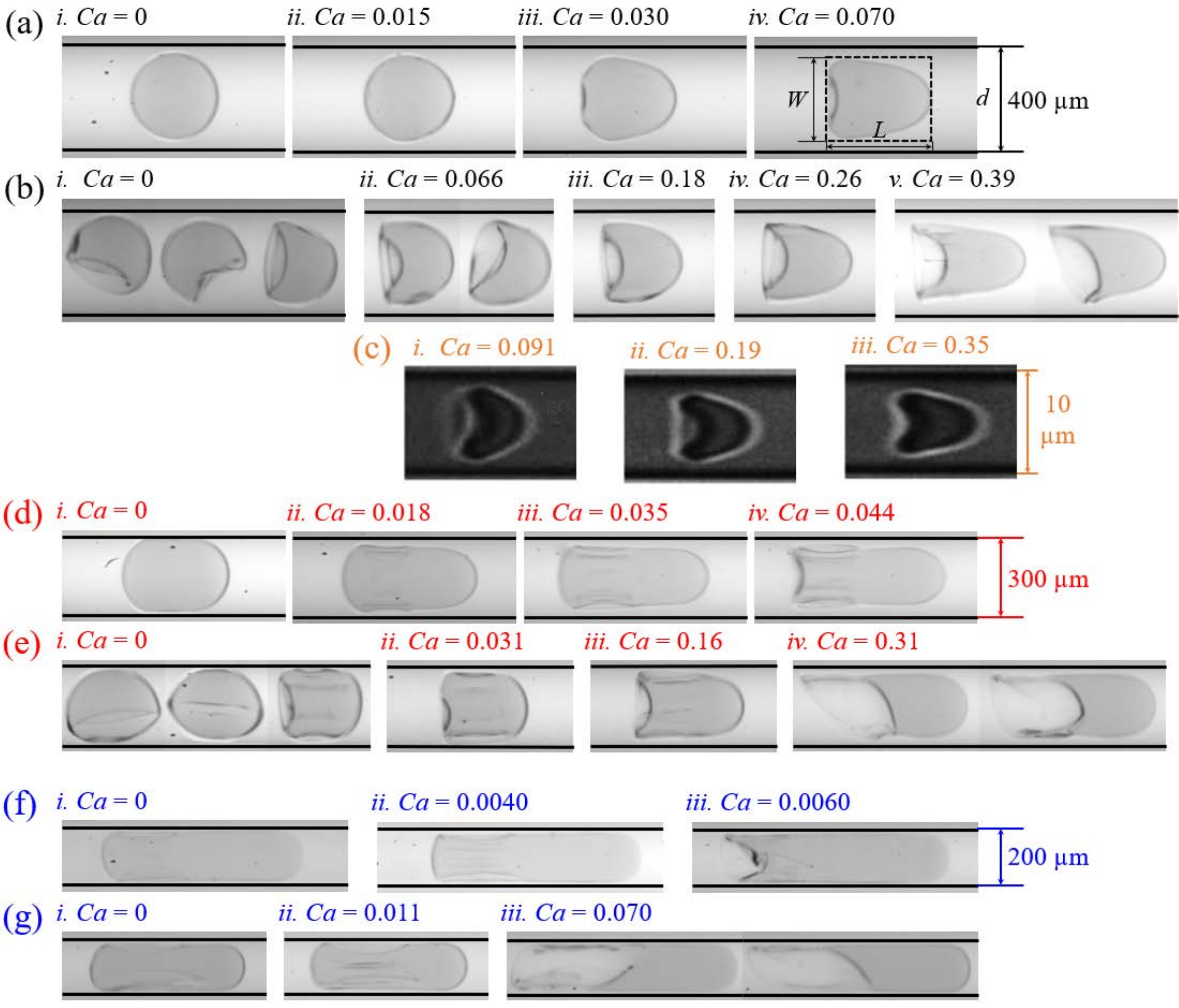}
    \caption{Steady shapes of spherical and deflated capsules and RBCs at different capillary numbers $Ca$ and confinement ratios $\beta$: (a) spherical capsules, $\beta$ = 0.875; (b) deflated capsules, $\beta$ = 0.755; (c) RBCs in a 10 $\upmu$m cylindrical silica capillary tube, $\beta$ $\approx$ 0.6 \cite{Lanotte_2014}; (d) spherical capsules, $\beta$ = 1.167; (e) deflated capsules, $\beta$ = 1.007; (f) spherical capsules, $\beta$ = 1.75; (g) deflated capsules, $\beta$ = 1.51. The flow direction is from left to right.}
    \label{fig:capsule deformation}
\end{figure*}

A comparison between the spherical and deflated capsule properties is listed in Table \ref{Table:capsule parameters}. The effective diameter $d_{\rm eff}$, which refers to the diameter of a sphere with the same volume as the capsules, is reduced from $d_{\rm o}$ = 350 $\upmu$m to approximately 302 $\upmu$m upon reduction of the capsule volume by 36\%. This means that the confinement parameter $\beta = d_{\rm eff}/D$ takes different values for spherical and deflated capsules in the same capillary tube. A 90:10 solution by volume of glycerol in water was used for both the suspending fluid and core of the spherical capsules, whereas pure glycerol was used for deflated capsules. Thus, there is no contrast between the viscosity of the capsule core and the suspending fluid.

Fig.~\ref{fig:capsule deformation} shows a comparison between steady state deformation of initially spherical capsules (Fig.~\ref{fig:capsule deformation}(a,d,f)) and deflated capsules (Fig.~\ref{fig:capsule deformation}(b,e,g)) in each of the three capillaries. For $D$ = 0.4 mm, both types of capsules are unconfined ($\beta$ = 0.875 for spherical capsules and $\beta$ = 0.755 for deflated capsules), so that they remain undeformed in the absence of flow; see panels (a-i) and (b-i) of Fig.~\ref{fig:capsule deformation}. The statically spherical capsule extends as $Ca$ increases, consistent with previous reports in the literature \cite{RISSO_2006}, and adopts a parachute shape seen in Fig.~\ref{fig:capsule deformation}(a-iii). We quantify capsule deformation as the ratio of the maximum length of the capsule to its maximum width $L/W$, which is measured on a closely fitted rectangular bounding box, enclosing the imaged capsule contour, parallel to the flow direction; see an example in panel (a-iv) of Fig.~\ref{fig:capsule deformation}. 

\begin{figure*}[h!]
    \centering
    \includegraphics[width=0.9\textwidth]{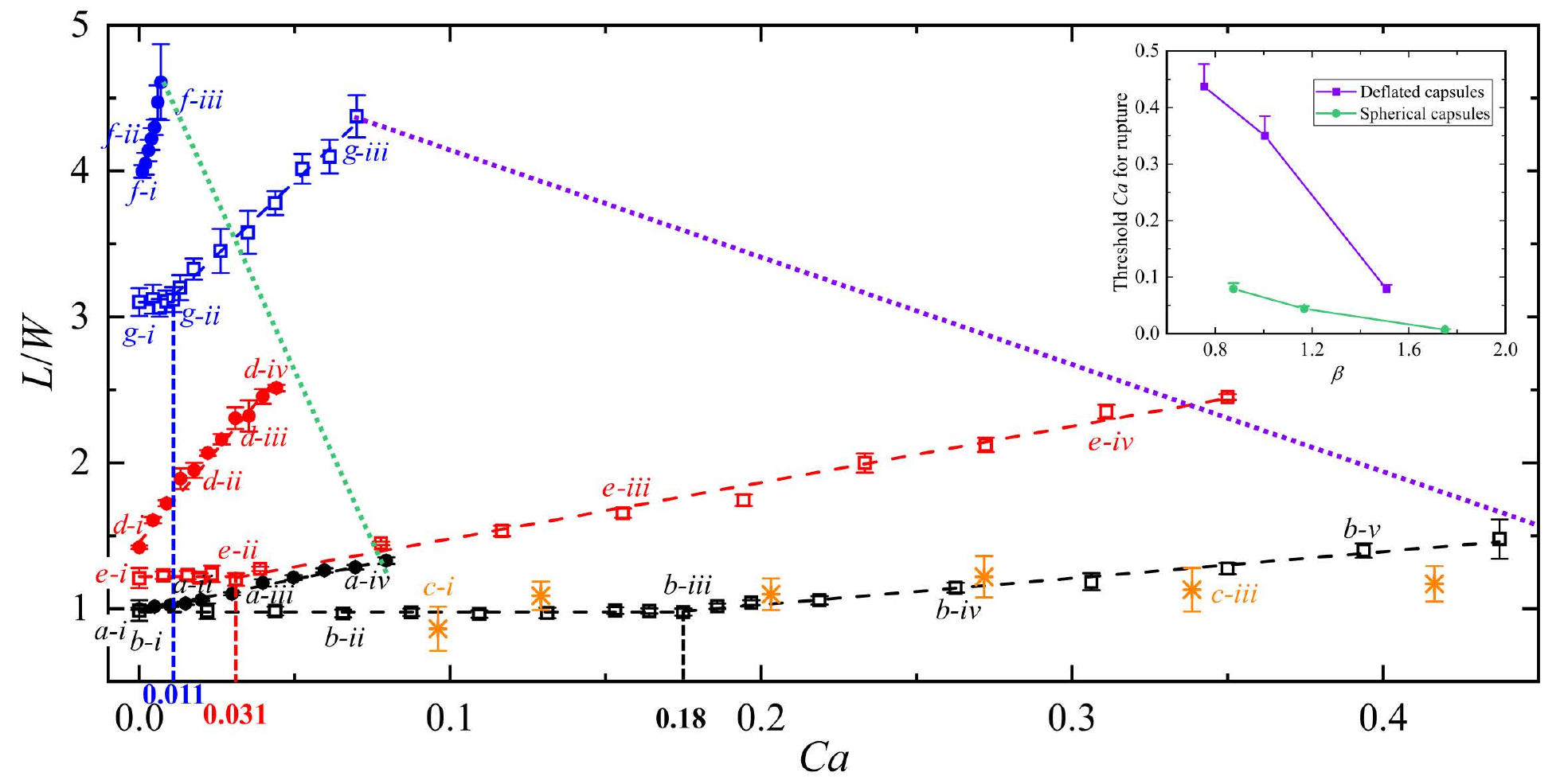}
    \caption{Variation of the capsule deformation ratio ($L$/$W$) with $Ca$. The solid dots and hollow squares correspond to the spherical and deflated capsules, respectively, where the different colour indicates different capillary diameters (black - 0.4 mm, red - 0.3 mm and blue - 0.2 mm). The error bar comes from the standard deviation of the parameters for at least 8 capsules. The dashed lines are the linear fittings of the data. The orange asterisks denote relative deformation of the RBCs in a 10 $\upmu$m silica capillary tube ($\beta$ $\approx$ 0.6) measured by Lanotte \emph{et al.} \cite{Lanotte_2014}. The green and purple dotted lines indicate the threshold $Ca$ for membrane rupture for spherical and deflated capsules, respectively, summarised in inset as a function of the confinement $\beta$. The labels refer to the capsule snapshots shown in Fig.~\ref{fig:capsule deformation}.}
    \label{fig:deformation ratio}
\end{figure*}

Fig.~\ref{fig:deformation ratio} shows that the deformation ratio $L/W$ varies approximately linearly as a function of $Ca$ for the initially spherical capsules (solid symbols) in all three capillaries. In contrast, the deflated capsules (open symbols) are buckled inward in the absence of confinement and flow (see Fig.~\ref{fig:capsule deformation}(b-i)), due to compressive stresses in the membrane induced by the volume reduction \cite{Quilliet_2012}. In this case, they adopt random orientations (Fig.~\ref{fig:capsule deformation}(b-i)), which results in a mean value of $L_0$/$W_0 \simeq 1$ at $Ca$ = 0 despite the fact that their volume is less than that of spherical capsules. In weak flow (Fig.~\ref{fig:capsule deformation}(b-ii)), these deflated capsules align with the shear flow so that their inward buckled region is situated at the rear and their shape resembles a parachute. However, they do not extend axially so that $L/W$ remains approximately constant up to a threshold value of the capillary number $Ca_{\rm th} = 0.18$ (Fig.~\ref{fig:capsule deformation}(b-iii)), where the shear forces on the capsule presumably balances the compressive pre-stress of the deflated capsule. Beyond this threshold, $L/W$ increases linearly with $Ca$ and the capsule retains its parachute-like shape (Fig.~\ref{fig:capsule deformation}(b-iv)). For high capillary numbers ($Ca \geq 0.32$), we observed both the symmetrical parachute-like and asymmetrical slipper-like shapes (Fig.~\ref{fig:capsule deformation}(b-v)). The elongated shapes of these deflated capsules are consistent with the deformation of RBCs in an unconfined capillary tube ($D = 10 \,\upmu$m, $\beta \approx 0.6$) at comparable capillary numbers $Ca$ \cite{Lanotte_2014}, as shown in panel (c) of Fig.~\ref{fig:capsule deformation}. The values of $\beta$ and $Ca$ of RBCs are calculated in the same way as for capsules, using an effective diameter of $3\, \upmu$m and a shear modulus of $5.4\,\upmu$N/m \cite{Tomaiuolo_2014}. RBCs (orange asterisks in Fig.~\ref{fig:deformation ratio}) exhibit deformation ratios that are marginally smaller than the deflated capsule data for $\beta = 0.755$ (black open squares in Fig.~\ref{fig:deformation ratio}) which is consistent with their smaller confinement parameter.

In capillaries with $D = 0.3$~mm and 0.2~mm, the spherical capsules are initially compressed into a cylindrical barrel shape with spherical end caps under tension (see panels (d-i) and (f-i) of Fig.~\ref{fig:capsule deformation}), so that the static deformation ratio $L_0$/$W_0$ exceeds unity ($Ca=0$). For increasing $Ca$, the capsules extend considerably in the axial direction while only marginally narrowing (Fig.~\ref{fig:capsule deformation}, panels (d-ii), (d-iii) and (f-ii)), so that the thickness of the lubrication layer between their surface and the wall of the tube remains approximately constant. We also observed axially oriented wrinkles around the circumference of the rear half of the capsule barrel (panels (d-ii), (d-iii) and (f-ii) of Fig.~\ref{fig:capsule deformation}) due to the significant lateral compression exerted on the capsule membrane, as discussed by Hu \emph{et al.} \cite{Hu_2011}. The capsule only adopts a parachute shape (Fig.~\ref{fig:capsule deformation}, (d-iv) and (f-iii)) for sufficiently large values of $Ca$ in excess of those required in the unconfined channel, because the confinement imposes axial tension on the capsule.

\begin{figure*}[h!]
    \centering
    \includegraphics[width=0.45\textwidth]{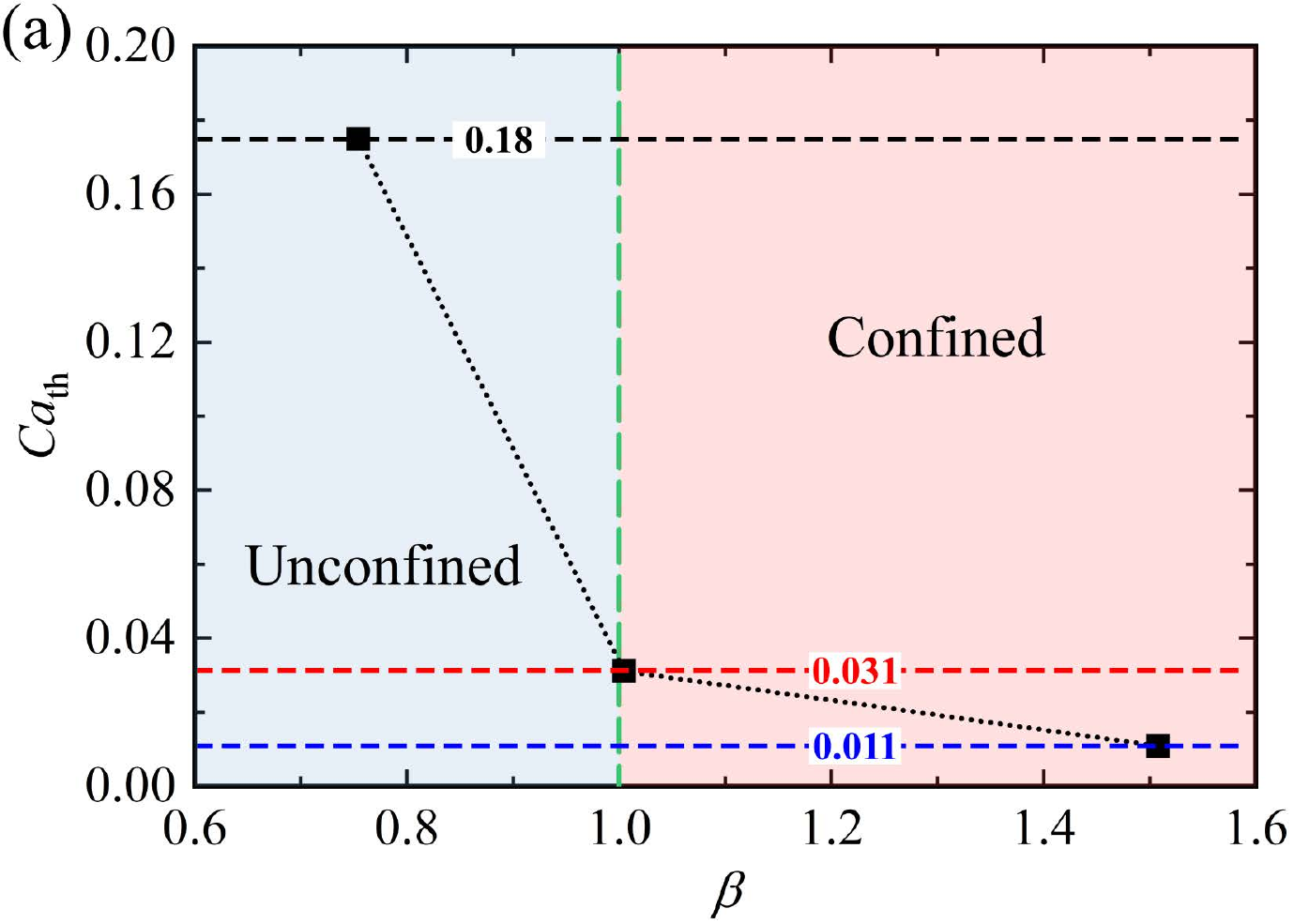}\includegraphics[width=0.45\textwidth]{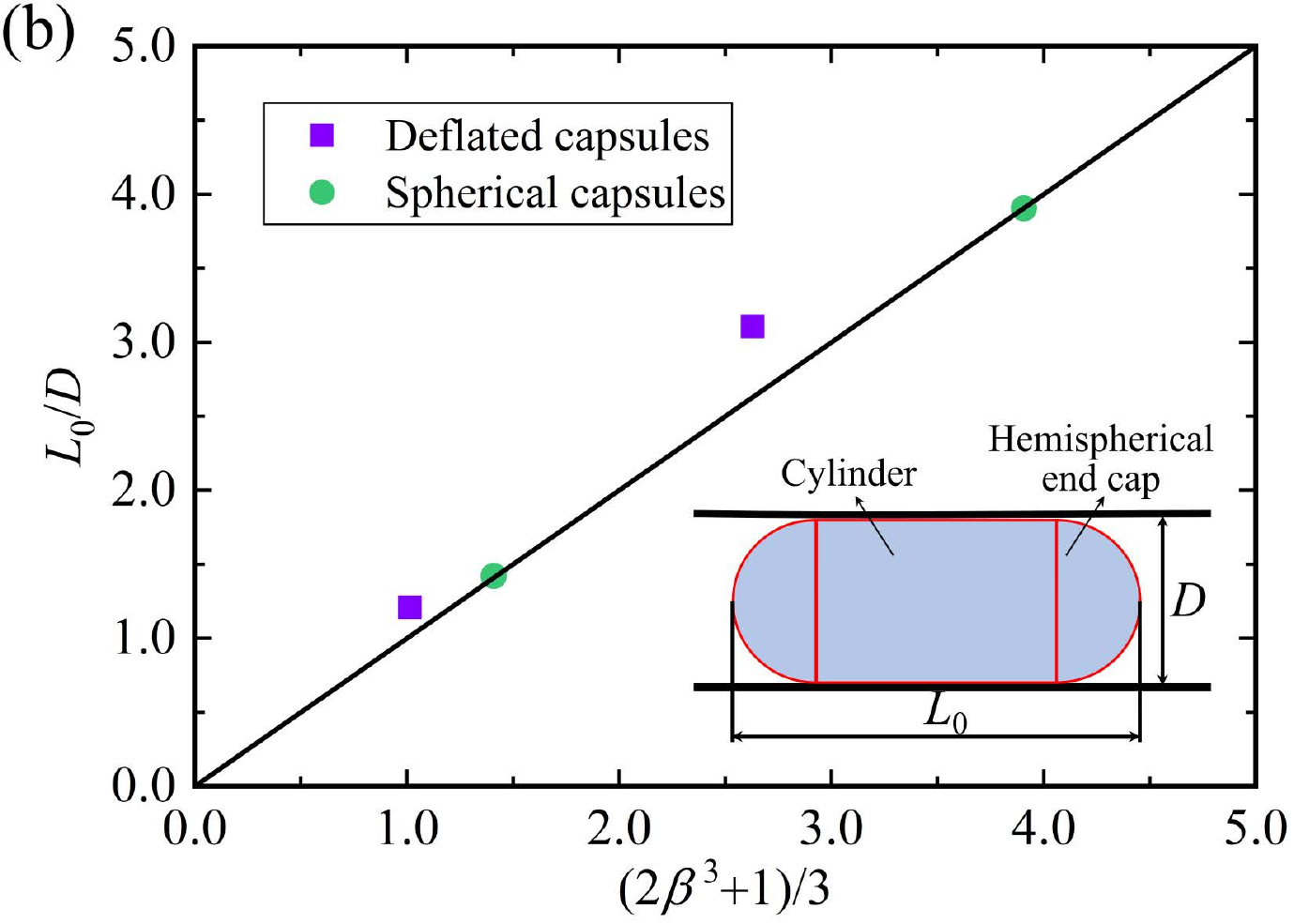} 
    \includegraphics[width=0.45\textwidth]{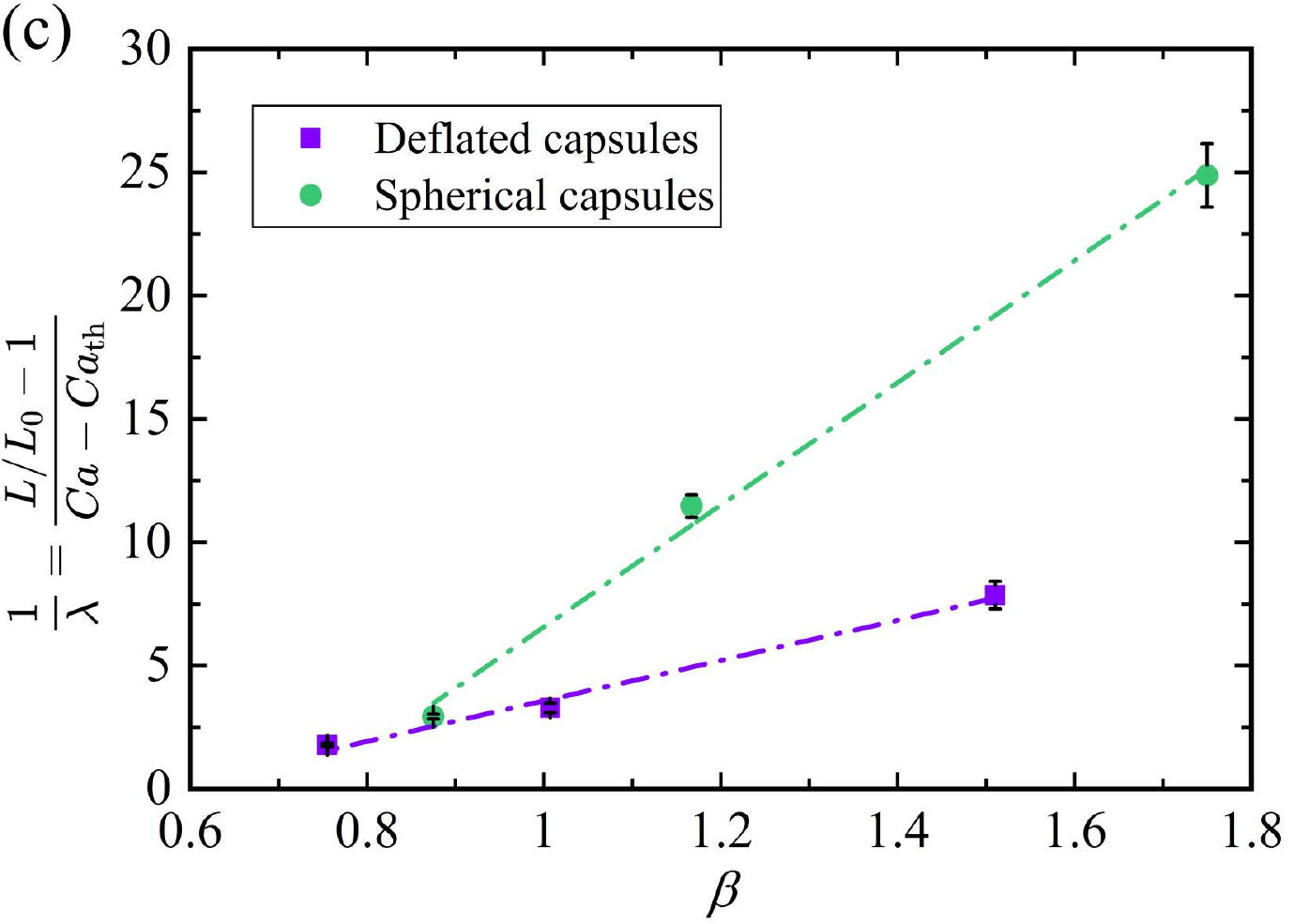}\includegraphics[width=0.45\textwidth]{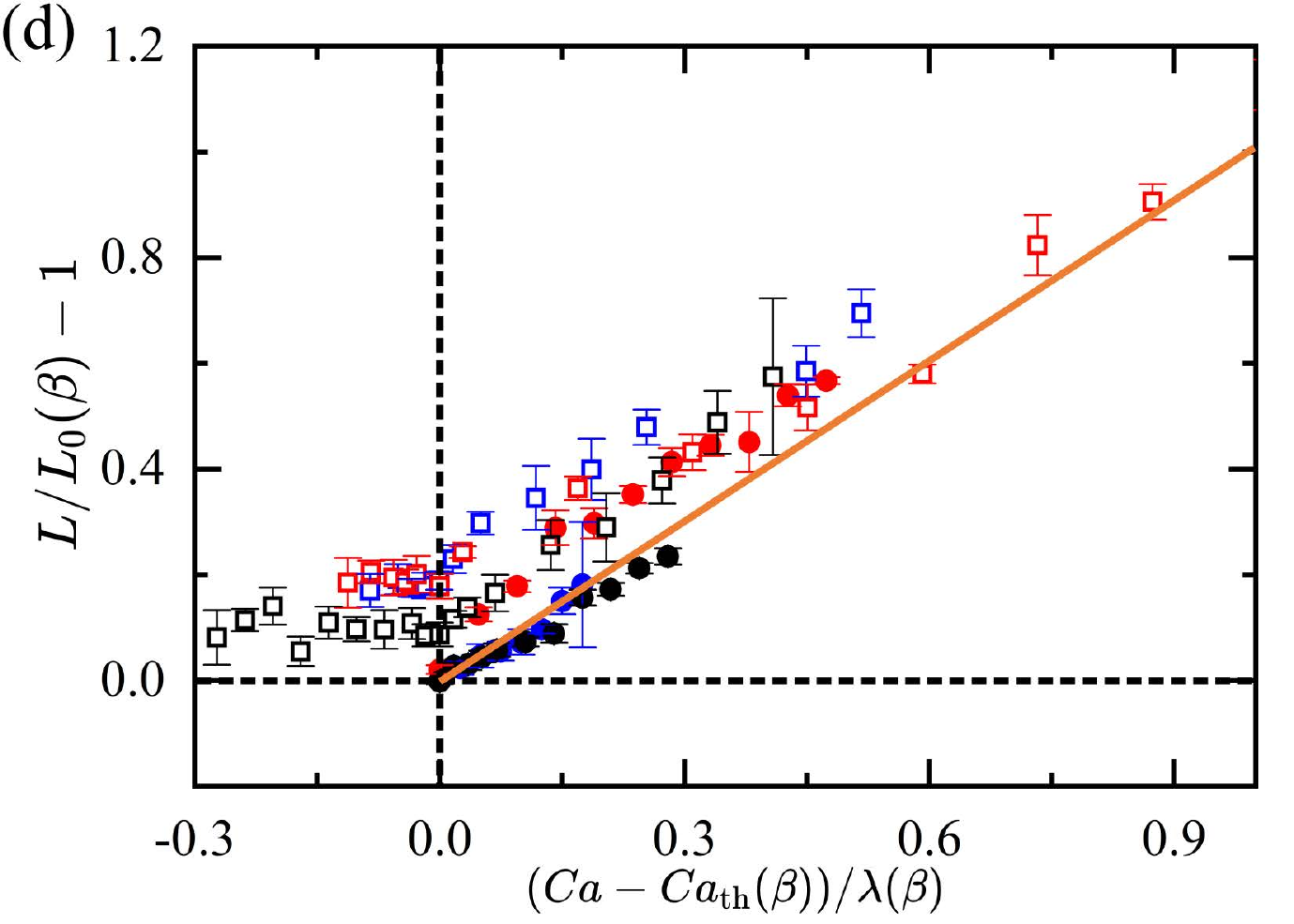}
    \caption{(a) The threshold value of $Ca$ for deflated capsules starting to be stretched as a function of $\beta$. The dotted line here is only used to guide the eyes. (b) The initial capsule length scaled by the tube diameter $L_0/D$ as a function of confinement $\beta$ for $\beta > 1$. The black line indicates the theoretical values predicted by Eq.~(\ref{eq:L_0/d to beta_all}). (c) The linear relations between the capsule elongation rate $1/\lambda$ and $\beta$. The dash dotted lines are the linear fittings and the error bars come from the errors of the fitting. (d) Variation of the scaled capsule elongation as a function of scaled $Ca$.}
    \label{fig:scaling}
\end{figure*}

In contrast, the deformation of deflated capsules with increasing $Ca$ is much closer to that observed in the absence of confinement (see Fig.~\ref{fig:capsule deformation}(b,e,g)). This is because the deflated capsule has excess membrane that can accommodate the increased confinement without significantly stretching its membrane axially. It follows that the threshold $Ca_{\rm th}$ beyond which the shear flow stretches the membrane decreases with increasing confinement $\beta$, as shown in Fig.~\ref{fig:scaling}(a), and we expect $Ca_{\rm th}$ to tend to zero for sufficiently high $\beta$. 
However, when $\beta$ decreases below unity, $Ca_{\rm th}$ increases sharply. 


Furthermore, we found that the spherical capsules typically break more easily under flow than the deflated capsules through rupture of their membrane. For spherical capsules, the threshold value of $Ca$ for membrane rupture is always less than 0.1, as shown with a green dotted line in the inset of Fig.~\ref{fig:deformation ratio}, where the vertical error bar indicates the $Ca$ interval between intact and ruptured capsules. This suggests that the spherical capsules do not adequately model the transport of RBCs, which can sustain higher relative shear stresses and typically remain intact for $Ca \geq O(10^{-1})$ \cite{Mendez_2019,Lanotte_2014}. In contrast, the deflated capsules deform within the same $Ca$ range as RBCs (see the orange asterisks in Fig.~\ref{fig:deformation ratio}) and only rupture at $Ca$ values of approximately five times the value for the spherical capsules (purple dotted line in Fig.~\ref{fig:deformation ratio}). Thus, the reduced volume and  excess membrane of the deflated capsules make them a useful proxy for the RBCs. 

The results of Fig.~\ref{fig:deformation ratio} indicate that the elongation of both spherical and deflated capsules is mainly determined by $Ca$ and $\beta$. In the absence of flow ($Ca = 0$), we use geometry to predict the dependence of the static capsule shape on $\beta$. If we assume both types of capsules are spherical for unconfined geometries ($\beta \le 1$), the initial capsule length and width approximately match their effective diameter $L_0=W_0=d_{\rm eff}$ and $L_0/D=\beta$. For $\beta >1$, the capsule shape can be approximated by a cylinder with hemispherical caps at both ends (see inset schematic in Fig.~\ref{fig:scaling}(b)), and volume conservation implies
\begin{equation}
    \frac{4}{3} \pi \left(\frac{d_{\rm eff}}{2}\right)^3 = \pi \left(\frac{D}{2}\right)^2\!(L_0 - D) + \frac{4}{3} \pi \left(\frac{D}{2}\right)^3,
\label{eq:cylinder-with-caps}
\end{equation}
which can be used to express $L_0/D$ in terms of the confinement parameter $\beta=d_\text{eff}/D$. Hence, Eq.~\eqref{eq:cylinder-with-caps} gives
\begin{equation}
    \frac{L_0}{D} (\beta) = 
    {\left\{\begin{matrix} 
  \qquad \beta, \qquad \quad \beta \le 1, \\  
  (2\beta^3+1)/3,  \quad \beta > 1.\label{eq:L_0/d to beta_all} 
\end{matrix}\right.}
\end{equation}
Fig.~\ref{fig:scaling}(b) shows that this expression accurately captures the experimentally measured values $L_0/D$ for $\beta > 1$ for spherical capsules (green symbols), while it slightly underestimates data for deflated capsules (purple symbols) owing to their buckled initial shapes. We use Eq.~\eqref{eq:L_0/d to beta_all} to rescale relative capsule elongation $L/D$ in Fig.~\ref{fig:scaling}(d), leading to the engineering strain $L/L_0 -1$ that is effectively zero at $Ca=0$ for the spherical capsules and rises slightly above zero for the deflated capsules at $Ca < Ca_{\rm th}$. 

Assuming that the cylindrical barrel region of a capsule is the primary source of its membrane surface extension in flow, and neglecting capsule width reduction ($W=D$) at increasing $Ca$, the relative surface extension is given by
\begin{equation}
    \frac{A-A_0}{A_0} \approx \frac{\pi D L- \pi D L_0}{\pi D L_0} = \frac{L}{L_0}-1\,,
\end{equation}
approximating the membrane surface strain with the capsule elongation strain $L/L_0-1$. Further assuming that local membrane deformations are sufficiently small for the total elongation strain to be proportional to the applied shear stress (under steady flow in a confined channel), we have\,\cite{Zhou_2022_2}  
\begin{equation}
    \frac{L}{L_0} - 1 \,\sim\, \frac{\mu_{\rm ext} \dot{\gamma}_{\rm w}}{G_{\rm s}} \frac{d_{\rm eff}}{2} \,\sim\, k\frac{d_{\rm eff}}{D} \frac{\mu_{\rm ext} \bar{u}}{G_{\rm s}} \,=\, k\,\beta\, Ca\,,
    \label{eq:stress balance}
\end{equation}
where $\dot{\gamma}_{\rm w}\sim \bar{u}/D$ approximates the wall shear rate, based on the mean velocity $\bar{u}$ and the channel diameter $D$, and $k$ is a dimensionless proportionality coefficient. The stress--strain balance given by Eq.~\eqref{eq:stress balance} captures the observed linear growth of the relative capsule elongation as a function of $Ca$ (Fig.~\ref{fig:deformation ratio}) and a linear relationship between the flow-induced elongation rate $(L/L_0-1)/Ca$ and the confinement parameter $\beta$ (Fig.~\ref{fig:scaling}(c)). 

Mendez and Abkarian \cite{Mendez_2019} rescaled the capillary number as $(\phi\,C)^{-1} \beta\, Ca$ to account for non-spherical capsule shapes and membrane pre-stress, introducing $\phi$, a geometric quantity characterising the capsule deflation (i.e., related to the reduced volume ratio $\alpha$), and $C$, a non-dimensional pre-stress constant. The inverse of the parameter $k$ in Eq.~\eqref{eq:stress balance} can therefore be interpreted as a measure of the combined effects of $\phi$ and $C$.  
Fig.~\ref{fig:scaling}(c) shows that the capsule elongation rate $1/\lambda = (L/L_0 - 1)/(Ca - Ca_{\rm th})$, where $Ca_{\rm th} = 0$ for spherical capsules, is approximately linear in $\beta$ over the considered experimental range for both spherical and deflated capsules, consistent with Eq.~\eqref{eq:stress balance}. 
A semi-empirical relation obtained by fitting the experimental data is shown in Fig.~\ref{fig:scaling}(c):
\begin{equation}
    \lambda^{-1}(\beta) = {\left\{\begin{matrix}
    24.77\beta - 18.22\,,\quad\;\; \text{for spherical capsules}\,,\\
    \;8.20\beta - 4.63\,,\qquad \text{for deflated capsules}\,. 
\end{matrix}\right.}
\label{eq:lambda to beta}
\end{equation}
The compressive stress associated with larger surface-to-volume ratio of the deflated capsules contributes to the smaller
slope coefficient in Eq.~\eqref{eq:lambda to beta} 
compared to the spherical capsules, which corresponds to the parameter $k$ in Eq.~\eqref{eq:stress balance}. 

Using Eq.~\eqref{eq:lambda to beta} to account for the confinement, we rescaled the excess capillary number above its threshold value, based on $Ca_{\rm th}$ (see Fig.~\ref{fig:scaling}(a)). Fig.~\ref{fig:scaling}(d) shows the scaled capsule elongation (engineering elongation strain) $L/L_0(\beta) - 1$ as a function of the effective capillary number $\left(Ca - Ca_\text{th}(\beta)\right)/\lambda(\beta)$. Therefore, in Fig.~\ref{fig:scaling}(d), both spherical and deflated capsules are unstrained at zero effective capillary number. This scaling is sufficient to approximately collapse the data onto a master curve (orange line in Fig.~\ref{fig:scaling}(d)), which indicates that the deflated capsules exhibit similar flow-induced deformation to the spherical ones for $Ca>Ca_\text{th}$. Fig.~\ref{fig:scaling}(d) also highlights that the deflated capsules can reach larger maximum elongation strains 
than the initially spherical capsules for a wide range of flow confinements.

\section{Conclusions and outlook}
\label{Conc}

In this paper, we have developed a 3D nested capillary microfluidic device which can robustly fabricate a large number of monodisperse PDMS microcapsules with ultra-thin and soft membranes as a physical model for RBCs. The geometrical and mechanical parameters of these non-ageing capsules (e.g., size, membrane thickness and membrane elasticity) are accurately controlled by varying the flow conditions and the chemistry of the membrane. This means that polydisperse suspensions of capsules of different size and/or stiffness can also be obtained by mixing populations of capsules manufactured under slightly different flow conditions. 

We deflate our capsules using osmosis to accurately match the reduced volume of real RBCs. This enables our capsules to exhibit the large elastic deformations characteristic of RBCs in confined flow without rupturing. We compare the steady propagation of initially spherical and deflated capsules for a wide range of capillary numbers and confinement ratios. The presence of compressive stresses induced by capsule deflation delays the elongation of deflated capsules for $Ca$ below a threshold value. Beyond this confinement-dependent threshold, capsule elongation increases approximately linearly with $Ca$. We show that only deflated capsules with the same reduced volume as RBCs exhibit comparable flow behaviour to RBCs over a similar range of capillary number ($Ca \geq O(10^{-1})$). Although both spherical and deflated capsules can adopt the typical parachute-like shape of steady RBC flow, symmetry-breaking into the slipper-like shape only occurs for the deflated capsules at sufficiently high $Ca$. To our knowledge, this is the first time when an experimental model quantitatively reproduces the steady-state deformations of RBCs.

Similar to computational models of RBCs, our capsules provide a physical model for the deformation of RBCs in flow. Such a model can be advantageous compared with experimentation on real RBCs because control and robustness enable the systematic variation of parameters. However, our capsules remain idealised in that they do not exhibit the biconcave shape of RBCs \cite{Wang_2012}, nor do they match the viscosity ratio between internal and external fluids and the encapsulating membrane is hyperelastic rather than viscoelastic \cite{Mendez_2019}. More generally, they  bypass key physiological effects, such as RBC-specific membrane biochemistry, cytoskeleton effects \cite{Mendez_2019} and aggregation phenomena \cite{Claver_a_2019}. 


Despite these simplifications, suspensions of these ultra-soft deflated capsules provide a powerful tool to explore the rheology of soft-particle suspension flows, with applications to haemodynamics and haemorheology in complex microvascular tissues, such as the human placenta \cite{Zhou_2022}, as well as to other areas of biomedicine and industry, such as targeted drug delivery \cite{Dinh_2020} and enhanced oil recovery \cite{Yiotis_2021}. 

\section*{Author contributions}
AJ and IC conceived this research; QC, NS and KS conducted the experiments; QC performed initial data analysis, with contribution from all the authors; QZ supplied reference RBC data; QC wrote the manuscript, which was edited by all the authors. 

\section*{Conflicts of interest}
There are no conflicts to declare.

\section*{Acknowledgements}
This work was supported by the UKRI EPSRC research grant (EP/T008725/1). QC acknowledges support by China Scholarship Council (grant No. 202006220020). QZ acknowledges support by the UKRI EPSRC grant (EP/T008806/1).



\balance

\bibliography{main.bib} 
\bibliographystyle{rsc} 

\clearpage

\appendix
\setcounter{page}{1}
\setcounter{section}{0}
\setcounter{figure}{0}
\setcounter{table}{0} 
\setcounter{equation}{0} 
\renewcommand{\thepage}{S\arabic{page}} 
\renewcommand{\thesection}{S\arabic{section}} 
\renewcommand{\thefigure}{S\arabic{figure}}
\renewcommand{\thetable}{S\arabic{table}}
\renewcommand{\theequation}{S\arabic{equation}}

\noindent{\Large \textbf{Electronic Supplementary Information (ESI)}}\label{sec:supplementary material}

\section*{S1 Structure design of the Teflon end caps}

\begin{figure} [h!]
    \centering
    \includegraphics[width=0.45\textwidth]{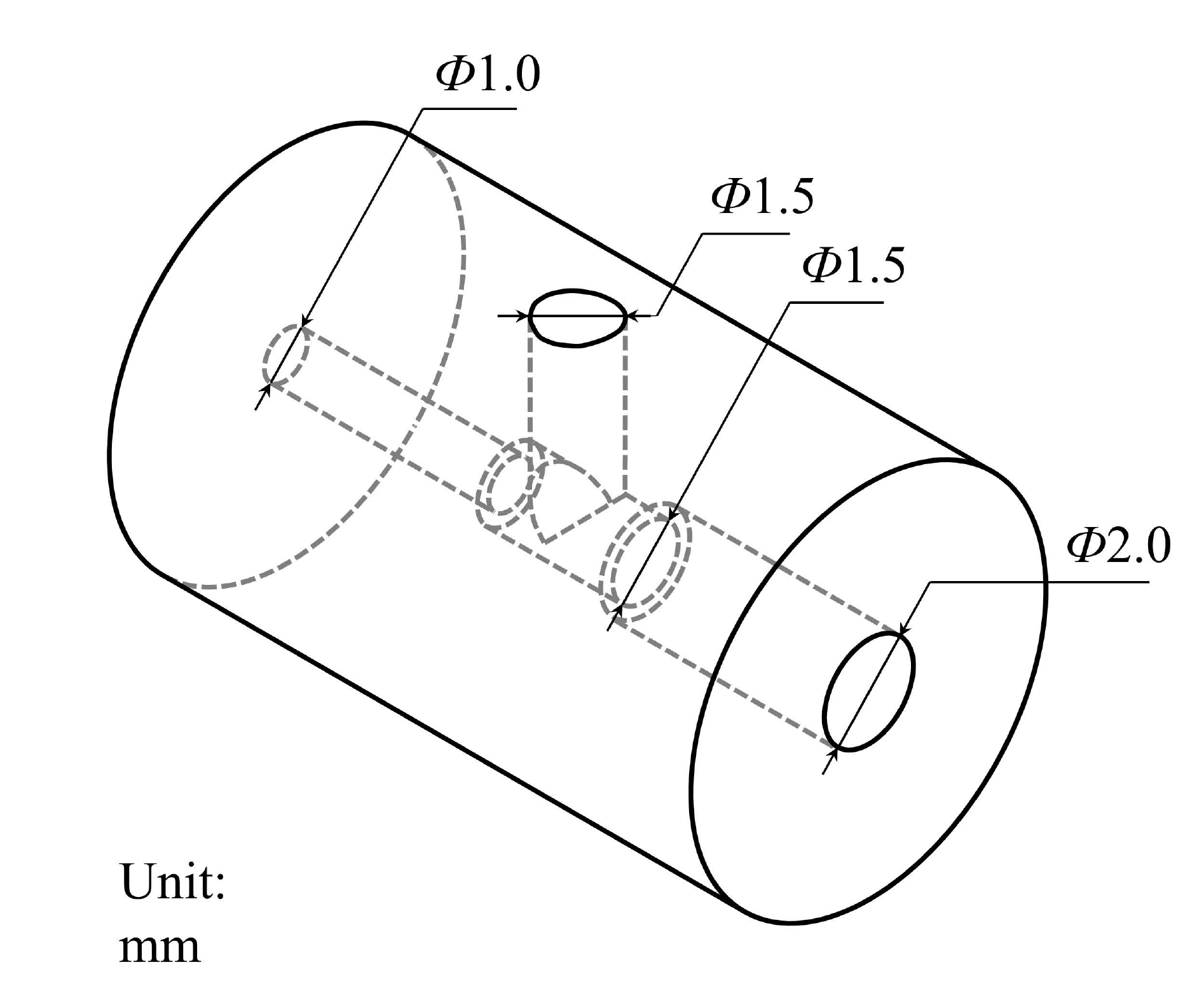}
    \caption{Structure of the Teflon connectors}
    \label{fig:Teflon connector}
\end{figure}

The Teflon end caps are fabricated according to the design reported by Levenstein \emph{et al.}$^1$, as shown in Fig. \ref{fig:Teflon connector}. Each of them has a large circular recess (2.0 mm) at the centre of one end of the Teflon cylinder. A smaller recess (1.5 mm) containing a hole (1.0 mm) through the centre of the cylinder is manufactured to allow the injection or collection capillary to insert into. A radial hole (1.5 mm) is drilled into the small recess to connect the tubing that supplies either the middle or outer liquid.

\section*{S2 Normal and abnormal flow behaviours in the device}

\noindent Video 1. Generation of double emulsions under normal conditions.

\noindent Video 2. Mixing of inner and outer phases due to the misalignments of injection and collection capillaries.

\noindent Video 3. Interface rupture due to the non-hydrophobic treatment of the injection capillary.

\noindent Video 4. Irregular interface duo to the non-hydrophilic treatment of the outer and collection capillaries.

\section*{S3 Measurement of Young's modulus}

\begin{figure*}[ht]
    \centering
    \includegraphics[width=0.3\textwidth]{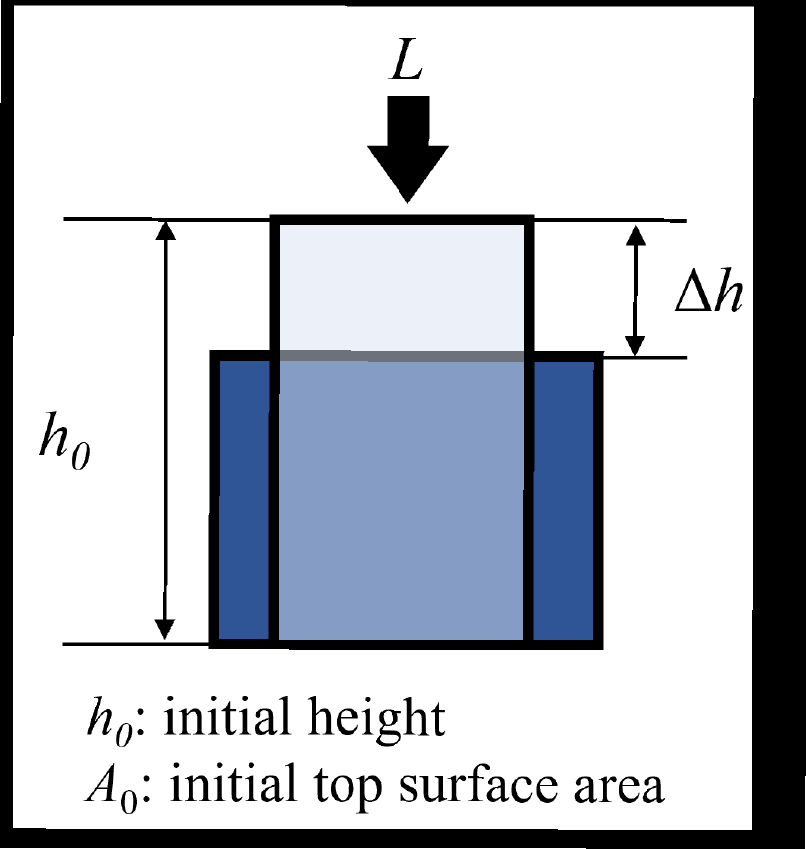}\includegraphics[width=0.5\textwidth]{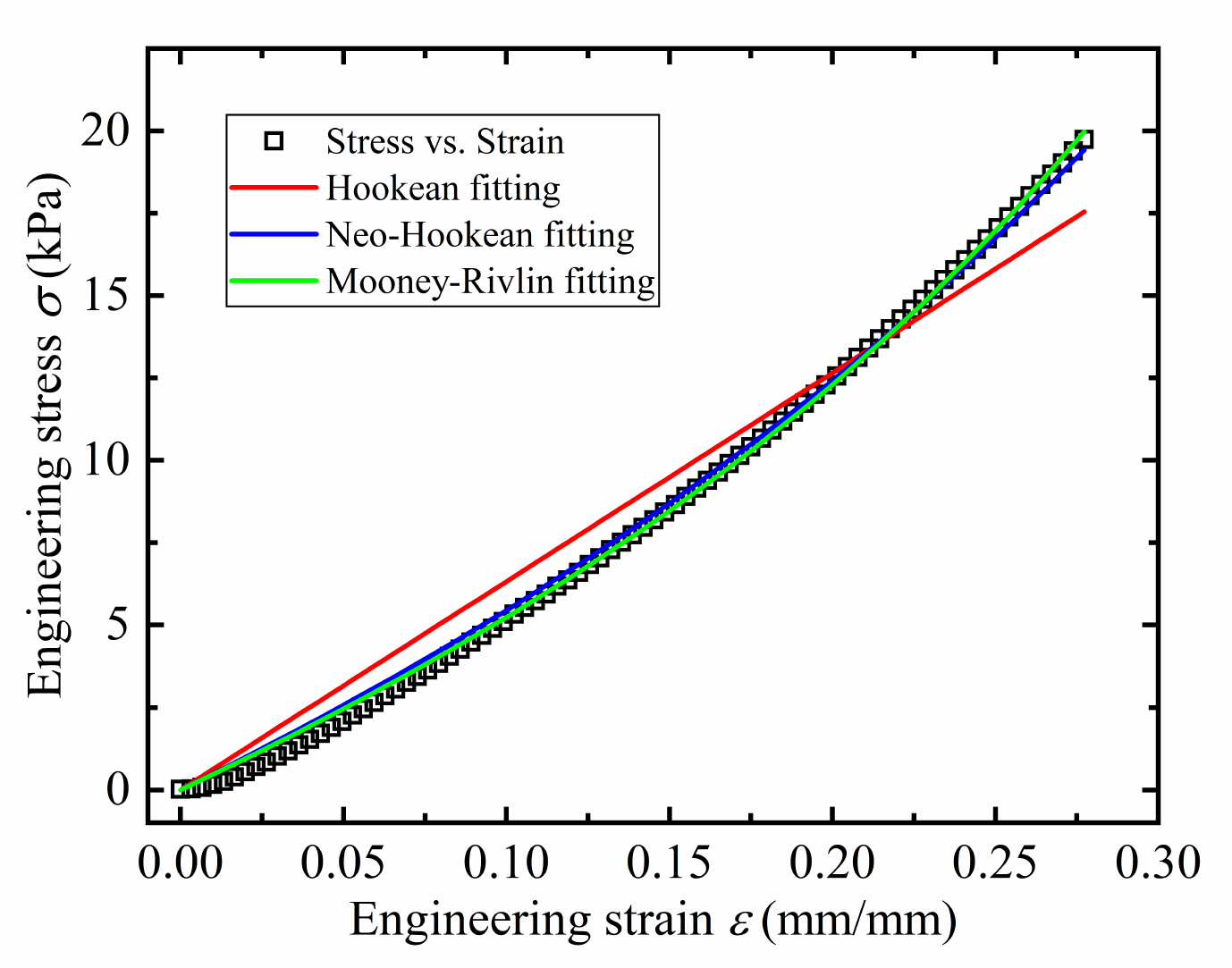}
    \caption{(Left) Schematic illustration of the compression test on a cylindrical sample. (Right) Fitting of three elasticity models, Hookean (red), neo-Hookean (black) and Mooney--Rivlin (green) models, to the stress-strain data (open squares). In this case, PDMS and corsslinker is mixed in 40:1, and the compression is performed at a rate of 0.01 mm/s.}
    \label{fig:compression}
\end{figure*}

We employ a commonly accepted compression method that has been described in the reference $^2$ to measure the Young's modulus of cured PDMS with the mixing ratio of PDMS base to the crosslinker ranging from 10:1 to 40:1. Liquid PDMS is moulded into cylindrical samples of 2.0 cm in diameter and 3.0 cm in height, followed by a uniaxial compression test with an Instron 5569 machine (Instron, High Wycombe, UK), where the load-displacement relationship is recorded, as shown in Fig. \ref{fig:compression} (Left). The measured load-displacement data is then converted into the nominal stress $\sigma$ (Pa) and train $\varepsilon$ (mm/mm) according to 

\begin{equation}
    \sigma = \frac{L}{A_0}
\end{equation}
and
\begin{equation}
    \varepsilon = \frac{\Delta h}{h_0},
\end{equation}
where $L$ (N) is the force exerted on the top surface of the test cylinder, $A_0$ is the initial area of the top surface, $\Delta h$ is the displacement of the top surface of the cylinder and $h_0$ is its initial height. We apply a thin lubricating layer of Vaseline to the top and bottom surfaces of the test cylinders to ensure they are deformed uniaxially. The Young's modulus is then obtained by fitting the experimental stress-strain data with the theoretical models of elastic materials. Here, we consider three common models: Hookean model, neo-Hookean model and the two-term Mooney-Rivlin model $^3$. By assuming the material to be isotropic, homogenrous and incompressible under uniaxial deformations, a relationship between the axial stress and one-dimensional strain is derived from the basic equations. Then, the nominal stress-strain relationships for these three models are expressed as below $^2$.\\
\noindent Hookean model:
\begin{equation}
    \sigma = E\varepsilon.  
    \label{eq:hookean}
\end{equation}

\noindent Neo-Hookean model:
\begin{equation}
    \sigma = 2C_1(1 - \varepsilon - \frac{1}{(1 - \varepsilon)^2}).
    \label{eq:neo-hookean}
\end{equation}

\noindent Mooney--Rivlin model:
\begin{equation}
    \sigma = (2C_1 + 4C_2((1 - \varepsilon)^2 + \frac{2}{1 - \varepsilon} - 3))(1 - \varepsilon - \frac{1}{(1 - \varepsilon)^2}).
    \label{eq:mooney-rivlin}
\end{equation}
In Eq. \ref{eq:hookean} to \ref{eq:mooney-rivlin}, $E$ (Pa) is the Young's modulus, $C_1$ (Pa) and $C_2$ (Pa) are the parameters obtained by fitting the experimental data with corresponding equations. For small deformations ($\varepsilon \ll 1$), the neo-Hookean and Mooney--Rivlin models reduce to
\begin{equation}
    \sigma = 6 C_1 \varepsilon.
\end{equation}
Then, the Young's modulus modulus is calculated as
 \begin{equation}
     E = 6C_1.
\end{equation}

\begin{table}[ht]
\centering
\caption{The Young's modulus $E$ obtained by fitting Hookean, neo-Hookean and Mooney-Rivlin models to the experimental stress-strain data (unit: kPa). The error comes from the standard deviation of results performed at different compression rates (from 0.01 mm/s to 1.00 mm/s).}
\begin{tabular}{llll} 
\hline
Mixing ratio & Hookean & Neo-Hookean & Mooney--Rivlin  \\ 
\hline
10:1         & 1633.8$\pm$32             & 1453.8$\pm$24                 & 1405.8$\pm$20                    \\
20:1         & 534.1$\pm$3.5             & 470.7$\pm$3.8                 & 430.3$\pm$9.8                    \\
30:1         & 148.1$\pm$1.2             & 133.9$\pm$1.3                 & 124.5$\pm$2.7                    \\
40:1         & 50.3$\pm$0.4             & 45.5$\pm$0.3                 & 41.6$\pm$0.7                    \\
\hline
\label{Table:Young's modulus}
\end{tabular}
\end{table}

Fig.~\ref{fig:compression} (Right) shows an example of the fittings of Eq. \ref{eq:hookean} (red), \ref{eq:neo-hookean} (blue) and \ref{eq:mooney-rivlin} (green) to the experimental stress-train data (black +). In this case, PDMS and its crosslinker are mixed in 40:1, the compression rate is 0.01 mm/s. The results show that Hookean model gives a poor fitting to the experimental data, which indicates that the generally accepted linear elasticity of PDMS under uniaxial compression is not suitable in our experiments. The two-term Mooney-Rivlin model provides the best prediction of the elastic behaviours of PDMS across the whole range of strain considered in this experiment. Table \ref{Table:Young's modulus} lists all the values of Young's modulus approximated by these three models, where the results predicted by Mooney--Rivlin model are considered for the capsule characterisation in this study.

\section*{References}
1  M. A. Levenstein, L. A. Bawazer, C. S. M. Nally, W. J. Marchant, X. Gong, F. C. Meldrum and N. Kapur, \textit{Microfluid. Nanofluidics}, 2016, \textbf{20}, 143.\\
2  S. Willshaw, \textit{On pattern-switching phenomena in complex elastic structures}, The University of Manchester (United Kingdom), 2012.\\
3  D. Barth{\`{e}}s-Biesel, A. Diaz and E. Dhenin, \textit{J. Fluid Mech.}, 2002, \textbf{460}, 211--222.

\end{document}